\documentclass{sig-alternate}
\usepackage{amssymb}

\usepackage{amsthm}

\usepackage{amsmath}
\usepackage{bbm}
\usepackage{balance}
\usepackage{graphicx}
\usepackage{epstopdf}
\usepackage{color}
\usepackage{algorithm}
\usepackage{algorithmic}
\usepackage{caption}
\usepackage{subcaption}

\usepackage{multirow}

\usepackage{enumitem}

\theoremstyle{plain}
\newtheorem{definition}{Definition}

\renewcommand{\vec}[1]{#1} 
\newcommand{\rv}[1]{\mathbf{#1}}    
\newcommand{\pr}[1]{\mathbb{P}\mathrm{r}\{#1\}} 
\newcommand{\pmatu}[3]{p^{#1}_{#2}({#3})}
\newcommand{\pvecu}[3]{\pi^{#1}_{#2}({#3})}
\newcommand{\exv}[2]{\mathbb{E}_{#2}\{#1\}} 
\newcommand{\ind}[1]{\mathbbm{1}_{#1}} 
\newcommand{\nn}{\nonumber}

\makeatletter
\def\@copyrightspace{\relax}
\makeatother

\begin{document}
	\setlength{\pdfpagewidth}{8.5in}
	\setlength{\pdfpageheight}{11in}

    \title{Privacy through Fake yet Semantically Real Traces}  
	
	\numberofauthors{2}
	\author{
		\alignauthor
			Vincent Bindschaedler\\
			\affaddr{UIUC}\\
		    \email{bindsch2@illinois.edu}
		\alignauthor
			Reza Shokri\\
			\affaddr{UT Austin}\\		    
			\email{shokri@cs.utexas.edu}
	}
    \maketitle

\begin{abstract}
Camouflaging data by generating fake information is a well-known obfuscation technique for protecting data privacy. The effectiveness of this technique in protecting users' privacy highly depends on the resemblance of fake information to reality, such that an adversary cannot easily filter such fake information out. In this paper, we focus on a very sensitive and increasingly exposed type of data: location data. There are two main scenarios in which fake traces are of extreme value to preserve location privacy: publishing datasets of location trajectories, and using location-based services. Despite advances in protecting (location) data privacy, there is no quantitative method to {\em evaluate} how realistic a synthetic trace is, and how much utility and privacy it provides in each scenario. Also, the lack of a methodology to {\em generate} privacy-preserving fake traces is evident. In this paper, we fill this gap and propose the first statistical metric and model to generate fake location traces such that both the utility of data and the privacy of users are preserved. 

We build upon the fact that, although geographically they visit distinct locations, people have strongly semantically similar mobility patterns, for example, their transition pattern across activities (e.g., working, driving, staying at home) is similar. 
We define a statistical metric and propose an algorithm that automatically discovers the hidden semantic similarities between locations from a bag of real location traces as seeds, without requiring any initial semantic annotations.  
We guarantee that fake traces are geographically dissimilar to their seeds, so they do not leak sensitive location information. We also protect contributors to seed traces against membership attacks.
Interleaving fake traces with mobile users' traces is a prominent location privacy defense mechanism. We quantitatively show the effectiveness of our methodology in protecting against localization inference attacks while preserving utility of sharing/publishing traces.
\end{abstract}

\section{Introduction}
\label{sec:intro}

Fake (dummy) information can protect privacy and security in many different systems such as web search \cite{HoweN09}, anonymous communications \cite{BertholdL03, DiazP04}, authentication systems \cite{JuelsR13}, and statistical analysis \cite{machanavajjhala2008privacy, rubin1993statistical}. In all these scenarios, the main challenge and the open problem is to generate context-dependent fake information that resembles genuine user-produced data and also provides an acceptable level of utility while enhancing privacy of users.

In this paper, we propose a systematic approach for preserving privacy of {\em location} data using fake traces. We focus on two practical scenarios: sharing location with location-based services, and publishing location datasets e.g., for research. In location-based systems, users hide their true location among fake location traces while connecting to a server to obtain contextual information about their whereabouts. This protects them against the location inference attacks. The benefit of the fake injection approach with respect to other obfuscation techniques, such as location perturbation \cite{AndresBCP13, HohGXA07, ShokriTTHB12}, 
is that it does not reduce the users' experienced service quality. Users only incur the overhead of filtering out the received information about fake locations. In publishing privacy-preserving location datasets, the purpose is to preserve the general statistics about human mobility. There is a utility loss associated with fake traces as they might not fully preserve all the characteristics of real traces. The challenge is to generate synthetic traces that semantically resemble the real traces yet do not leak information about the exact geographic locations visited by any particular individual. This gives rise to a tradeoff between utility and privacy that is inherent to privacy-preserving systems. 

There has been some preliminary work on using fake location queries to protect users' location privacy \cite{ChowG09, KidoYS05, Krumm09b, ShokriTDHL11, YouPL07}. However, they are based on very simple heuristics such as i.i.d. location sampling, sampling locations from a random walk on a grid with uniform probability, and using road trip algorithms to generate driving traces between two random locations. In this paper, we quantitatively show that these methods fail to protect location privacy against inference attacks. Besides, what these methods are missing are (i) a {\em metric} that captures how realistic a synthetic location trace is with respect to human mobility, so it cannot be easily detected by attacker, (ii) a {\em generative model} that produces samples of synthetic yet realistic traces according to such a metric, while preserving utility and ensuring that the synthetic traces do not themselves leak information about any individual. In this paper, we present the first formal methodology to solve these problems and to generate fake yet semantically real location traces for protecting location privacy. We also enforce, and quantitatively measure, privacy against location privacy attacks.

Our scheme is based on the fact that mobility patterns of different individuals share some semantic features, regardless of which geographic locations they visit. These common features of human mobility stem from their similar lifestyles. The mobility patterns share a similar structure that reflects the general behavior of a population (even at a high level \cite{song2010limits}).  
We model the mobility of each individual in two dimensions: {\em geographic} and {\em semantic}. The geographic features are mostly specific to each individual (hence are sensitive), whereas the semantic features are mostly generic and representative of human mobility behavior (hence are useful). We extract the common semantic features of mobility patterns and use them to generate realistic synthetic traces, without leaking the geographic features of any individual's locations.

Consequently, we define two metrics to quantify the similarity between human mobility models: The geographic similarity metric between two individuals captures how correctly we can predict locations of one knowing the mobility model of the other one. This metric helps us to capture the spatiotemporal information leakage of a fake trace about a real trace. The semantic similarity metric reflects how well two traces match in terms of their semantic features. %
We assume that we have a dataset of real traces. We develop an algorithm that automatically learns the semantic correlation between locations and transitions between locations.\footnote{Note that we do not annotate the locations (as home, work, ...), nor we use any annotation as an input to compute the similarity between locations. We only rely on location traces as input.} 
To generate semantically realistic fake traces, we transform a (real) seed trace into the semantic domain and probabilistically sample (fake) location traces that are consistent with it. Thus, a generated sample resembles a typical sequence of locations that could have been visited by some real individual. We then design a {\em rejection sampling} assertion to ensure that a fake trace does not leak information about the locations in its seed trace, i.e., we reject those who do not meet the privacy requirements. Thereby, we protect privacy of seed traces against the following threats: {\em Inference} attacks (to learn which locations the seed contributors have visited), and {\em membership inclusion} attack (to learn if a particular individual with certain semantic habits has been in the seed dataset). If a fake trace's geographic similarity or its intersection with its real seed trace exceeds a given threshold, we reject it and sample a new trace. We also ensure that the semantic similarity between fake and seed traces cannot be used against anonymity of seed traces. To this end, we reject the fake trace if there is no $k-1$ alternative real traces that could have been the seed for generating the fake (i.e., the differential semantic similarity with the fake trace is below a threshold). This additionally guarantees a {\em plausible deniability} for each seed trace. The resulting pool of fake traces can later be drawn upon for use by e.g., users' smart-phones. The fake traces generated from the seed database can be used to protect location privacy of any user (not the contributed of the seed database). The privacy tests guarantee that we preserve privacy of seed traces. Additionally, we show that the generated fakes can significantly protect privacy of LBS users against inference attacks.

{\em Our Contributions}: In summary, the novelty of this work is twofold. (1) We introduce the notion of semantic similarity for mobility patterns, we propose both a metric for it, and an algorithm to quantify the semantic similarity between location traces. We also automatically learn the semantic relation between locations. (2) We propose a generative model for fake location traces that are semantically similar to real traces. We also guarantee plausible deniability to individuals whose real traces are used as seed in our algorithm. Our software tool, given a set of real location traces, generates fake traces based on our theoretical framework. We run our algorithms on a real-world dataset collected by Nokia \cite{kiukkonen2010towards}. We show the effectiveness of our fake traces in protecting privacy of users in two main scenarios: location-based services and published location datasets, while preserving utility.

\section{Our Scheme}\label{sec:solution}

\begin{figure}[t]
\centering
\includegraphics[width=1.0\columnwidth]{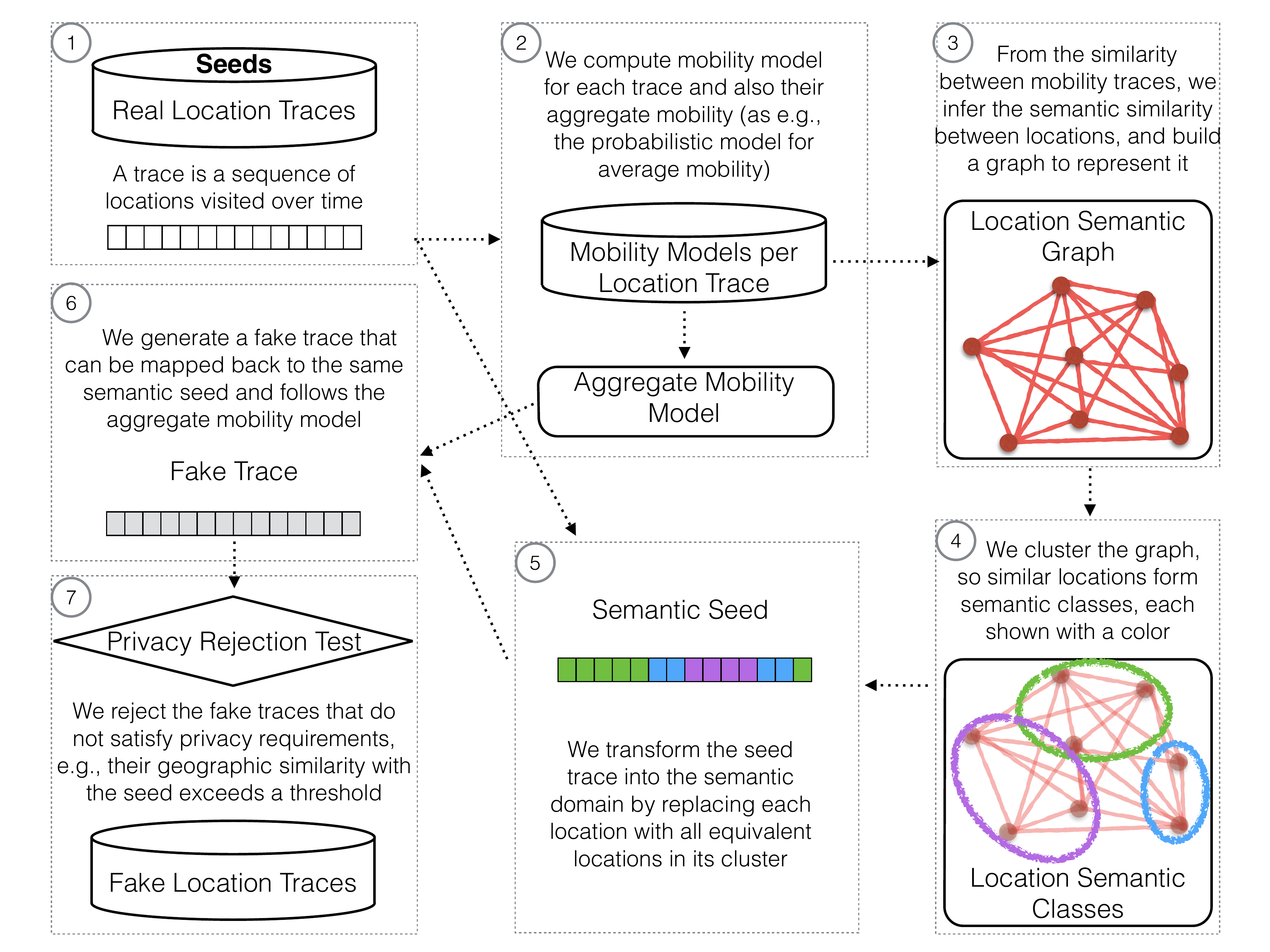}%
\caption{Sketch of the proposed scheme.}%
\label{fig:sketch}%
\end{figure}

In this section, we present a sketch of, and describe the main intuition behind, our scheme for generating fake traces. We assume that time and space are discrete, so a location trace is represented as a sequence of visited locations over time. In our scheme, we generate a fake trace through a multiple step process. Figure~\ref{fig:sketch} illustrates the details.

\subsection{Computing the Semantic Similarity} \label{sec:solution:similarity}

The first step is to compute the semantic similarity between the set of locations in an area. We learn these similarity values automatically from a {\em dataset} of real location traces. For each trace (i.e., the sequence of locations visited by a single individual) in the dataset, we first compute a probabilistic {\em mobility model} that represents the visiting probability to each location and transition probability among the locations (see Section~\ref{sec:similarity:mobility}). The mobility model encompasses the spatiotemporal behavior of each individual with respect to different locations. Time, duration, and probability (frequency) of visiting a location, as well as the probable comings from and goings to locations are all computable from the mobility model. So, it implicitly reflects the types of activities that an individual might carry out in each location (and over a sequence of locations).

We analyze and discover the semantic relation between different locations in a consistent manner by considering all locations together. To this end, we propose a {\em semantic similarity metric} (see Section~\ref{sec:similarity:semantic}). Intuitively, we assign a higher similarity value to the pair of locations at which different individuals have similar spatiotemporal activities. Thus, our metric tries to find the optimal way to map the visited locations in a pair of traces such that the mapping maximizes the statistical similarity between their mobility models. The semantic similarity metric is therefore the statistical similarity between mobility models under the optimal mapping between locations. This means that if we were to translate the locations of this pair according to the discovered best mapping, they would follow the same mobility model when their semantic similarity is high. For example, consider Alice and Bob spending all day at their respective work locations $w_A$ and $w_B$, and all night at their respective home locations $h_A$ and $h_B$. Obviously, their mobility models are semantically very similar, although it might be the case that $h_A \neq h_B$ and $w_A \neq w_B$. In this example, the best semantic mapping between locations will be $w_A \leftrightarrow w_B$ and $h_A \leftrightarrow h_B$.

For each pair of mobility models for traces in our dataset, we compute their semantic similarity as well as the best semantic mapping between their locations. We then aggregate all the location matchings across all trace pairs, with weights based on the semantic similarity between mobility models, and construct a {\em location semantic graph}, where the nodes are locations and the weight of the edges is the average semantic similarity between the locations over the dataset.

\subsection{Forming Location Semantic Classes} \label{sec:solution:classes}

The location semantic graph enables us to find what locations have similar meanings for different people, so they have similar activities in those places. The locations that have higher semantic similarity can be grouped together to represent one {\em location semantic class}. To this end, we run a clustering algorithm on the location semantic graph to partition locations into distinct classes. Locations that fall into the same class are visited in the same way by different people regardless of their geographic positions.\footnote{\scriptsize Their visit probability, time of visit, and the probabilities of transition from/to them to/from other locations of the same type is similar.} Thus, we can consider them as being semantically equivalent. So, using the notations of our previous example, $w_A$ and $w_B$ should belong to the same cluster that can represent ``workplace'' locations, and $h_A$ and $h_B$ should be grouped into another cluster representing residential or ``home'' locations.

\subsection{Generating a Fake Trace} \label{sec:solution:generate}

We use the location semantic classes as the basis to generate fake traces. In addition to being semantically realistic, the fake traces must be geographically consistent with the general mobility of individuals in the considered area. For example, the speed of moving in some locations differs depending on the time, or the probabilities of taking different paths is different. To capture these patterns, we compute an {\em aggregate mobility} model from the traces in our dataset. We can, for example, compute this by averaging the mobility models that we constructed on the traces.

The goal is to generate fake traces that are semantically similar to real traces. In order to construct a fake trace, our algorithm starts with a {\em seed} trace and converts it to a probabilistically generated semantically similar synthetic trace which is consistent with the aggregate mobility model. We pick the seed trace at random from the trace dataset. The seed trace, similar to other location traces in the database, is composed of a sequence of geographic locations. In our algorithm, we first {\em transform} the geographic seed trace into the semantic domain, then we use the transformed semantic trace to {\em sample} from the state space of all geographic traces that could have been transformed to the same semantic trace. The transformation and sampling procedures, which are at the heart of this step, are done as follows.

For the seed trace {\em transformation}, we replace the geographic locations in the seed with the locations that are in the same semantic class. This seed {\em semantic trace} is a sequence of location sets. For the fake trace {\em sampling}, we address the following problem. We want to construct a trace that follows the aggregate mobility model under the constraint that its locations over time are subset of locations of the seed semantic trace. Hence, both the fake trace and the seed trace can be transformed to the same semantic trace. We add some randomness to the locations in the semantic trace to allow higher number of possible fake traces. Many algorithms can be used to sample the fake trace that satisfies our constraints. We make use of dynamic programming algorithms that construct the traces efficiently (Section~\ref{sec:sampling}).

We can repeatedly generate fake traces from each seed trace in the dataset, each of which having a probability according to the aggregate mobility model. After generating each fake trace, however, we need to make sure that it is not geographically similar to the seed trace. This is because we do not want to leak information about the real seed trace. To this end, we add a test to compute the {\em geographic similarity} (see Section~\ref{sec:similarity:geographic}) between the seed trace and the fake trace to {\em reject} the sample traces that are more similar than a threshold to the seed trace. Thus, we make sure that the semantically similar synthetic traces are indeed geographically dissimilar to the traces in our dataset, hence not leaking information about visited locations in the real traces.

\begin{table}[t]%
\begin{center}
{\scriptsize
\setlength{\tabcolsep}{.3\tabcolsep}
\begin{tabular}{c|l}
  \hline
  $\mathcal{R}$ & Set of locations\\[0.07cm]
  $R$ & Number of locations \\[0.07cm]
  $r$ & A location \\[0.07cm]
  $\mathbf{r}$ & Random variable associated with a location \\[0.07cm]
  $T$ & Number of time periods \\[0.07cm]
  $\tau$ & A time period \\[0.07cm]
  $p(u)$ & Transition probability matrix of user $u$ \\[0.07cm]
  $\pi(u)$ & Visiting probability vector of user $u$ \\[0.07cm]
  $\langle p(u), \pi(u) \rangle$ & Mobility profile of user $u$ \\[0.07cm]
  $p^{\mathbf{x}}_{\mathbf{y}}(u)$ & Probability of $\mathbf{x}$ given $\mathbf{y}$ according to $u$'s mobility model\\[0.07cm]
  $d(\cdot)$ & A distance function (between locations) \\[0.07cm]
  $M_d(p, q)$ & Mallows distance between probability distributions \\& $p$ and $q$ based on a distance function $d(\cdot)$ \\[0.07cm]
  $\sigma$ & A permutation function \\[0.07cm]
  $\mathsf{sim_G}(u, v)$ & Geographic similarity between mobility of $u$ and $v$ \\[0.07cm]
  $\mathsf{sim_S}(u, v)$ & Semantic similarity between mobility of $u$ and $v$ \\[0.07cm]
  $\sigma_u^v$ & Optimal semantic mapping between locations of $u$ and $v$ \\[0.07cm]
  $\mathcal{S}$ & Set of real traces used as seeds to generate fake traces\\[0.07cm]
  $\langle \bar{p}, \bar{\pi} \rangle$ & Aggregate mobility model\\[0.07cm]
  $\mathcal{C}$ & A partition on $\mathcal{R}$, representing location semantic classes.\\& $\mathcal{C}_i$ is the set of locations in class (partition) $i$ \\[0.07cm]
  $\mathcal{F}$ & A set of fake locations generated from $\mathcal{S}$ \\[0.07cm]
  \hline
\end{tabular}
}
\end{center}
\caption{Table of notations}\label{tab:notation}%
\end{table}%

\section{Mobility Similarity Metrics}\label{sec:similarity}

In this section, we present a probabilistic model for mobility, and propose two metrics to analyze the geographic and semantic similarity between two mobility models. Table~\ref{tab:notation} presents the list of notations that we use in this paper.

\subsection{Mobility Model}\label{sec:similarity:mobility}

We model the user mobility as a time-dependent first-order Markov chain on the set of regions (locations). As users have different behavior and mobility patterns during different periods of time, we assume that time is partitioned into time periods, e.g., morning - afternoon - evening - night. So, the mobility profile $\langle p(u), \pi(u) \rangle$ of a given user $u$ is a {\em transition} probability matrix of the Markov chain associated with the user's mobility (from a region to another), and the user's {\em visiting} probability distribution over the regions, respectively. Note that these probabilities are dependent on each other, and together they constitute the joint probability of two regions that are subsequently visited by the user. The entry $p^{r'}_{r, \tau, \tau'}(u)$ of $p(u)$ is the probability that user $u$ will move to region~$r'$ in the next time instant (which will be in time period~$\tau'$), given that she is now (in time period~$\tau$) in region~$r$. The entry $\pi^{r}_{\tau}(u)$ is the probability that user $u$ is in region $r$ in time period $\tau$. Let the random variable $\rv{A}^{t}_u$ represent the actual location of user $u$ at time $t$, and $\tau^{t}$ be the time period associated with $\rv{A}^{t}_u$. So, the mobility profile of a given user $u$ consists of the following probabilities:
\begin{align}\label{eq:p_and_pi}
&p^{r'}_{r, \tau, \tau'}(u) = \pr{ \rv{A}^{t+1}_u = r' \,|\, \rv{A}^{t}_u = r ; \tau^{t+1} = \tau', \tau^{t} = \tau }, \nn\\
&\pi^{r}_{\tau}(u) = \pr{ \rv{A}^{t}_u = r ; \tau^{t} = \tau }
\end{align}

This Markovian model can predict the location of an individual to a great extent, as it takes both location and time aspect into account. It can become even more precise, by increasing its order, or by enriching its state. We can, for example, include multiple granularities of locations, and model the mobility of a user on the set of e.g., pair of (location, neighborhood) in addition to the time dimension. Our framework can incorporate all these new dimensions similar to the way we model the time periods. To learn the probabilities of the mobility profile \eqref{eq:p_and_pi}, from location traces, we can use maximum likelihood estimation (if the traces are complete) or make use of algorithms such as Gibbs sampling (if the traces have missing locations or are noisy) \cite{ShokriTLH11}.

\subsection{Mobility Similarity Metrics}

We propose two metrics to compare the mobility of two users and compute their similarities: {\em geographic} and {\em semantic} similarity. In this subsection, we describe the intuition behind these metrics, and in the following subsections, we formally define and provide the algorithms to compute them.

The {\em geographic similarity} metric captures the correlation between location traces that are generated by two mobility profiles. It reflects if two users visit similar locations over time with similar probabilities and if they move between those locations also with similar probabilities. Using this metric, for example, two individuals who live in the same region A and their workplace is in the same region B potentially have very similar mobilities, as they spend their work hours in B and most of their free time in A. 

There are very few people whose mobility patterns have a high geographic similarity with each other. However, if we ignore the exact locations that are visited by different people, we observe that they share similar patterns for visiting locations with similar semantics (locations therein they have similar activities). For example, most people visit and stay at a single location from each evening until its subsequent morning. These locations differ from one individual to another, but have the same semantic for them: home.

One can imagine the semantic dimension of locations as a coloring on the locations in a map. Instead of computing the correlation between location traces at the geographic level, we can also compute such correlation at the semantic level (by reducing the set of locations to the set of colors and computing the similarity of color traces). This is the intuition behind our {\em semantic similarity} metric. In this case, if the pair of locations that two individuals visit over time have the same semantic, their mobility models are also semantically similar (even if they are in two different cities, i.e., have no geographic similarity). Hence, in this example, if we transform trace A by replacing its locations with their corresponding semantically similar locations in trace B, the transformed trace becomes geographically similar to B. So, two traces are semantically similar if their locations can be mapped to each other that accordingly one trace can be transformed to a geographically similar trace to the other. 

\subsection{Geographic Similarity Metric}\label{sec:similarity:geographic}

We define this similarity metric based on the Earth Mover's Distance (EMD) for probability distributions. The EMD is widely used in a range of applications including image processing~\cite{Rubner98, Rubner00}. The EMD can be understood by thinking of the two distributions as piles of dirt. In this interpretation, the EMD represents the minimum amount of work needed to turn one pile of dirt (i.e., one distribution) into the other; the cost of moving dirt being proportional to both the amount of dirt and the distance to the destination. The special case of EMD for probability distributions has been shown to be equivalent to the Mallows distance~\cite{Levina01}.

Let $\rv{X}$ and $\rv{Y}$ be discrete random variables with probability distributions $\vec{p}$ and $\vec{q}$, such that $\pr{\rv{X} = x_i} = p_i$ and $\pr{\rv{Y} = y_i} = q_i$, respectively, for $i = 1, 2, \ldots, n$. We also have $\sum_i p_i = 1$ and $\sum_i q_i = 1$.

\begin{definition} (From~\cite{Levina01})
    Let $d(\cdot)$ be an arbitrary distance function between $\rv{X}$ and $\rv{Y}$. The \emph{Mallows distance} $M_d(\vec{p}, \vec{q})$ is defined as the minimum expected distance between $\rv{X}$ and $\rv{Y}$ with respect to $d(\cdot)$ and to any joint distribution function $\vec{f}$ for $(\rv{X}, \rv{Y})$ such that $\vec{p}$ and $\vec{q}$ are the marginal distributions of $\rv{X}$ and $\rv{Y}$, respectively.
    \begin{equation}
        M_d(\vec{p}, \vec{q}) \! := \! \min_{\vec{f}} \! \left\{ \exv{d(\! \rv{X},\! \rv{Y} \!)}{\vec{f}} \! : \! (\! \rv{X}, \! \rv{Y} \!) \! \sim \! \vec{f},\! \rv{X} \! \sim \! \vec{p}, \! \rv{Y} \! \sim \! \vec{q} \right\} ,
    	\label{eq:emdprob}
    \end{equation}
    where the expectation, minimized under $\vec{f}$, is
    \begin{equation}
        \exv{d(X, Y)}{\vec{f}} = \sum_{i = 1}^n \sum_{j = 1}^n f_{ij} \  d(x_i, y_j).
    \end{equation}
\end{definition}

In addition to the constraints $\sum_{i = 1}^n \sum_{j = 1}^n f_{ij} = 1$ and $f_{ij} \geq 0$, for all $i$, $j$, the joint probability distribution function $\vec{f}$ must also satisfy $\sum_{i = 1}^n f_{ij} = q_j$ and $\sum_{j = 1}^n f_{ij} = p_i$.

Note that, for given $\vec{p}$ and $\vec{q}$, the minimum $\vec{f}$ is easily computed by expressing the optimization problem as a linear program.

Using the previous definition, we define the geographic similarity metric based on the Mallows distance.

\begin{definition}
    Let $d(\cdot)$ be an arbitrary distance function. The \emph{dissimilarity} between two mobility profiles $\langle p(u), \pi(u) \rangle$ and $\langle p(v), \pi(v) \rangle$ (belonging to individuals $u$ and $v$), is defined as the expected Mallows distance of the next random locations $\mathbf{r'}$ and $\mathbf{r''}$ according to the mobility profiles of $u$ and $v$, respectively. More formally, it is
    \begin{equation}
        \exv{M_d(\pmatu{\mathbf{r'}}{\mathbf{r}, \mathbf{\tau}, \mathbf{\tau'}}{u}, \pmatu{\mathbf{r''}}{\mathbf{r}, \mathbf{\tau}, \mathbf{\tau'}}{v}) }{(u)},
        \label{eq:ds1}
    \end{equation}
    where $\pmatu{\mathbf{r'}}{\vec{r}, \vec{\tau}, \vec{\tau'} }{u}$ and $\pmatu{\mathbf{r''}}{\vec{r}, \vec{\tau}, \vec{\tau'}}{v}$ denote the conditional probability distributions of the next location, given the current location and the current and next time periods. The Mallows function is computed over random variables $\mathbf{r'}$ and $\mathbf{r''}$, and the expectation is computed over random variable $\mathbf{r}$ and time periods $\tau$ and $\tau'$.

    We define the {\em geographic similarity} between mobility patterns of $u$ and $v$ as
    \begin{equation}
        \mathsf{sim_G}(u, v) := 1 - \exv{M_d(\pmatu{\mathbf{r'}}{\mathbf{r}, \vec{\tau}, \vec{\tau'}}{u}, \pmatu{\mathbf{r''}}{\mathbf{r}, \vec{\tau}, \vec{\tau'}}{v}) }{}.
    \end{equation}
    \label{def:ds1}
\end{definition}

We compute the geographic {\em dissimilarity} using the law of total expectation. This also clarifies its meaning by showing more directly the role of the random variables.
\begin{align}
    & \exv{M_d(\pmatu{\mathbf{r'}}{\mathbf{r}, \vec{\tau}, \vec{\tau'}}{u}, \pmatu{\mathbf{r''}}{\mathbf{r}, \vec{\tau}, \vec{\tau'}}{v}) }{}  \nn \\
    &= \sum_{r, \tau, \tau'}  M_d{(\pmatu{\mathbf{r'}}{r, \tau, \tau'}{u}, \pmatu{\mathbf{r''}}{r, \tau, \tau'}{v})} \cdot \pmatu{r, \tau, \tau'}{}{u}.
    \label{eq:ds1ex}
\end{align}

This is simply the average, for each time and location, of the EMD between the distributions of the next location of $u$ and $v$. So, for each current location (and time), we use the EMD to compute the dissimilarity between the distributions representing the next locations of users $u$ and $v$, respectively. The current location is taken according to user $u$'s mobility profile, making this definition asymmetric.

For a particular distance function $d(\cdot)$, the Mallows distance definition can be expanded and previous expressions can be further simplified. This is the case for $d(i, j) := \ind{i \neq j}$, for which $M_d(\vec{p}, \vec{q})$, for arbitrary probability distributions $\vec{p}$ and $\vec{q}$, has closed form $1 - \sum_i{\min{\{p_i, q_j\}}}$.

Using the dissimilarity metric, we can compute the {\em geographic similarity} between the mobility profiles $\langle p(u), \pi(u) \rangle$ and $\langle p(v), \pi(v) \rangle$, for any distance function (e.g., hamming distance, Euclidean distance). For example, considering hamming distance $d(r, r') = \ind{r \neq r'}$, the geographic similarity is:
\begin{align}
    &1 - \sum_{r, \tau, \tau'} \left( 1 - \sum_{r'} \min \{ {\pmatu{r'}{r, \tau, \tau'}{u}, \pmatu{r'}{r, \tau, \tau'}{v} } \} \right) \cdot \pmatu{r, \tau, \tau'}{}{u} \nn \\
    &= \sum_{r, r', \tau, \tau'} \pmatu{\tau'}{r, \tau}{u} \pvecu{r, \tau}{}{u} \min \{ {\pmatu{r'}{r, \tau, \tau'}{u}, \pmatu{r'}{r, \tau, \tau'}{v} } \}.
    \label{eq:as1}
\end{align}

We emphasize that this definition leads to an asymmetrical similarity measure, i.e. the similarity of $u$ to $v$ need not be the same as the similarity of $v$ to $u$. In principle, this metric can also be computed using measures other than EMD. For example, one can use Kullback-Leibler divergence measure \cite{CoverT94} to compute the difference between two probability distributions, ignoring the distance between the locations. We emphasize that we use EMD, in our geographic similarity metric, as we also want to include the distance function $d(\cdot)$ between locations while computing the difference between two distributions (i.e., mobility models).

Consider now the computation of the geographic similarity. For the case, $d(r, r') = \ind{r \neq r'}$, the computation according to closed-form of \eqref{eq:as1} takes $O(T^2 \cdot R^2)$ operations, where $T$ is the number of time periods and $R$ is the number of locations (regions). For arbitrary $d(\cdot)$ with no closed-form expressions, the geographic similarity is obtained through $T^2 \cdot R$ EMD computations. Each of these EMD computations involves minimizing the Mallows distance, that is equivalent to solving the linear program given by \eqref{eq:emdprob}.

\subsection{Semantic Similarity Metric}\label{sec:similarity:semantic}

The semantic similarity metric builds upon the basic assumption that for two individuals $u$ and $v$ there exists an (unknown) semantics mapping $\sigma$ of locations $\mathcal{R}$ onto itself (i.e. a permutation) such that $\mathcal{R}$, for $u$ and $\sigma^{-1}(\mathcal{R})$, for $v$ semantically match. It is important to note that assuming such a mapping does not commit us to trying to learn it based on modeling location semantics directly. Instead, we define the (hidden) semantic similarity between $u$ and $v$ as the maximum geographic similarity taken over all possible mappings $\sigma$. We define semantic similarity metric as follows.

\begin{definition}

    Let $\sigma$ be the mapping of locations of $u$ to locations of $v$. Let $\mathbf{r}$, $\mathbf{r'}$, and $\mathbf{r''}$ be random variables for locations, and $\tau$ and $\tau'$ be two time periods. We define the semantic {\em dissimilarity} between $u$ and $v$ for moving in the sequence of time periods $\{\tau, \tau'\}$ as
    \begin{align}
        \mathcal{D}_u^v(\{\tau, \tau'\}) := \min_\sigma \mathbb{E} \left[ M_d ( p^{\mathbf{r'}}_{\mathbf{r}, \tau, \tau'}(u), p^{\sigma(\mathbf{r''})}_{\sigma(\mathbf{r}), \tau, \tau'}(v) ) \right],
        \label{eq:hdkp}
    \end{align}
    where the Mallows distance $M_d(\cdot)$ is computed over the random variable $\mathbf{r'}$ and the expectation is computed over the random variable $\mathbf{r}$ given time periods $\tau$ and $\tau'$.

    Now, we define the {\em semantic similarity} between $u$ and $v$ over any sequences of time periods as
    \begin{equation}
        \mathsf{sim_S}(u, v) := 1 - \mathbb{E} \left[ \mathcal{D}_u^v(\{\tau, \tau'\}) \right].
        \label{eq:hdk}
    \end{equation}
\end{definition}

What we compute in \eqref{eq:hdkp} is the minimum geographic mobility dissimilarity between $u$ and $v$ where the locations of $v$ are relabeled and mapped to locations of $u$ according to the permutation function $\sigma_u^v$ (which is the $\sigma$ that minimizes \ref{eq:hdkp}). The intuition is the following. Consider two individuals $u$ and $v$ are at $\mathbf{r}$ and $\sigma(\mathbf{r})$, respectively, at time period $\tau$. The Mallows distance $M_d$ computes how dissimilar their movement will be to the next location which are represented with random variables $\mathbf{r'}$ for $u$ and $\sigma(\mathbf{r''})$ for $v$. If, according to a mapping, the way that they move between these locations is similar, they behave similarly with respect to those locations. If this is true across different time periods and location pairs, their mobilities are similar. So, the semantic similarity between two individuals is determined by $\sigma_u^v$.

We compute this metric at two different levels of accuracy of the mobility model. If we only consider the visiting probability $\pi$ part of each individual's mobility profile, we compute $\mathbf{sim_S}$ as follows. Let us consider the hamming distance function $d(r, r') = 1_{r \neq \sigma^{-1}{(r')}}$. In this case, we can compute the semantic similarity metric as
\begin{equation}
    1 - \sum_{\tau} \pr{\tau} \ \max_\sigma \sum_r \min\{\pi_{\tau}^r(u), \pi_{\tau}^{\sigma(r)}(v)\} .
    \label{eq:hs0}
\end{equation}

Note that the computation of \eqref{eq:hs0} requires finding the mapping $\sigma$ which maximizes the inner term for each time period $\tau$. Since there are $R!$ possible candidates for the maximum mapping $\sigma$, a brute-force approach is inefficient. However, the problem's structure resembles that of a linear assignment. Focusing on the inner sum, we see that each term (each $r$) can be associated with $R$ values of $\sigma(r)$ independently of the other components of $\sigma$. To recast the problem as a linear assignment, we construct a bipartite graph where the nodes represent $\mathcal{R}$ and $\mathcal{\sigma{(R)}}$, and each edge represents the association (through $\sigma$) of $r$ with $\sigma(r)$. The maximum weight assignment of the constructed bipartite graph gives the permutation $\sigma$. The running time of this procedure is $O(T \cdot R^3)$ using the Hungarian algorithm~\cite{munkres1957algorithms}.

We compute the semantic similarity for the case where we consider the more accurate mobility profile $\langle p, \pi \rangle$ as follows.
\begin{equation}
    1 - \sum_{\tau, \tau'} \ \max_\sigma \sum_{r, r'} \pi^{r, \tau}(u) p_{r, \tau}^{\tau'}(u) \min\{p_{r, \tau, \tau'}^{r'}(u), p_{\sigma(r), \tau, \tau'}^{\sigma(r')}(u)\}
    \label{eq:hs1}
\end{equation}

It is not known whether there is an efficient algorithm to compute the semantic similarity according to \eqref{eq:hs1}. 
The difficulty comes from having to consider assignments of pairs: $(r, r')$ to $(\sigma(r), \sigma(r'))$, which makes this computation resemble the Quadratic Assignment Problem (QAP) \cite{pardalos1994quadratic},
 known to be NP-Hard and APX-Hard.
 The semantic similarity \eqref{eq:hs1} can nevertheless be computed through approximation techniques such as Simulated Annealing~\cite{brooks1995optimization}, or the Metropolis-Hastings algorithm~\cite{metropolis1953equation}.
Nevertheless, \eqref{eq:hs1} can be approximated using techniques such as Simulated Annealing~\cite{brooks1995optimization}, or the Metropolis-Hastings algorithm.~\cite{metropolis1953equation}.
 We use this algorithm to compute the semantic similarity metric, in the case of considering both visiting and transition probabilities of the individuals' mobility models (see~\cite{mackay2003information} for details). The idea is to find good approximations to the quantity of interest (for us $\sigma$) through probabilistic local exploration of the solution space. At each step, we replace our current permutation with a new solution randomly selected from its neighbors (e.g. a permutation which differs in two positions). The output of the algorithm is the best permutation found so far when the algorithm terminates (after some fixed number of iterations). It is known that the starting permutation can have an impact on the quality of the output. In our case, we expect the permutation found during the computation of \eqref{eq:hs0} to be a good starting point.

\section{Generating Fake Traces}\label{sec:sampling}

\begin{figure}[t!]
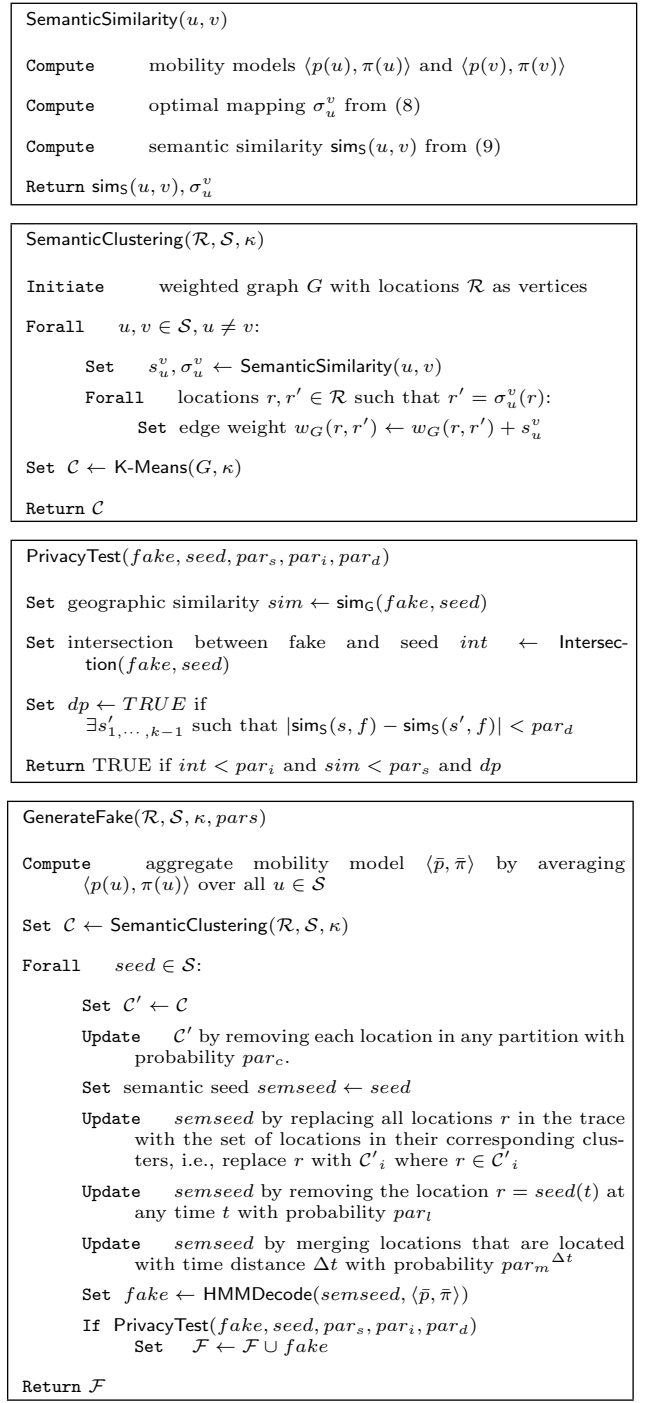

\centering
{\scriptsize
\fbox{
\parbox[c]{0.95\linewidth}{
{\scriptsize
\textsf{SemanticSimilarity}$(u, v)$\\[-0.2cm]
\begin{description}
	\item[\tt Compute] \qquad mobility models $\langle p(u), \pi(u) \rangle$ and $\langle p(v), \pi(v) \rangle$
	\item[\tt Compute] \qquad optimal mapping $\sigma_u^v$ from \eqref{eq:hdkp}
	\item[\tt Compute] \qquad semantic similarity $\mathsf{sim_S}(u, v)$ from \eqref{eq:hdk}
\end{description}
\texttt{Return} $\mathsf{sim_S}(u, v), \sigma_u^v$
}}}\\[0.2cm]
\fbox{
\parbox[c]{0.95\linewidth}{
{\scriptsize
\textsf{SemanticClustering}$(\mathcal{R}, \mathcal{S}, \kappa)$\\[-0.2cm]
\begin{description}
	\item[\tt Initiate] \qquad weighted graph $G$ with locations $\mathcal{R}$ as vertices
	\item[\tt Forall] \quad $u, v \in \mathcal{S}, u \neq v$:
		\begin{description}
			\item[\tt Set] \quad $s_u^v, \sigma_u^v$ $\leftarrow$ \textsf{SemanticSimilarity}$(u, v)$
			\item[\tt Forall] \quad locations $r, r' \in \mathcal{R}$ such that $r'=\sigma_u^v(r)$:
				\begin{description} 
					\item[\tt Set] edge weight $w_G(r, r') \leftarrow w_G(r, r') + s_u^v$
				\end{description}
		\end{description}
	\item[\tt Set] $\mathcal{C} \leftarrow$ \textsf{K-Means}$(G, \kappa)$
\end{description}
\texttt{Return} $\mathcal{C}$
}}}\\[0.2cm]
\fbox{
\parbox[c]{0.95\linewidth}{
{\scriptsize
\textsf{PrivacyTest}$(fake, seed, par_s, par_i, par_d)$\\[-0.2cm]
\begin{description}
	\item[\tt Set] geographic similarity $sim \leftarrow \mathsf{sim_G}(fake, seed)$
	\item[\tt Set] intersection between fake and seed $int \leftarrow$ \textsf{Intersection}$(fake, seed)$
	\item[\tt Set] $dp \leftarrow TRUE$ if\\$\exists s_{1,\cdots,k-1}'$ such that $|\mathsf{sim_S}(s, f) - \mathsf{sim_S}(s', f)| < par_d$
\end{description}
\texttt{Return} TRUE if $int < par_i$ and $sim < par_s$ and $dp$
}}}\\[0.2cm]
\fbox{
\parbox[c]{0.95\linewidth}{
{\scriptsize
\textsf{GenerateFake}$(\mathcal{R}, \mathcal{S}, \kappa, pars)$\\[-0.2cm]
\begin{description}
	\item[\tt Compute] \qquad aggregate mobility model $\langle \bar{p}, \bar{\pi} \rangle$ by averaging $\langle p(u), \pi(u) \rangle$ over all $u \in \mathcal{S}$
	\item[\tt Set] $\mathcal{C} \leftarrow$ \textsf{SemanticClustering}$(\mathcal{R}, \mathcal{S}, \kappa)$
	\item[\tt Forall] \quad $seed \in \mathcal{S}$:
		\begin{description}
			\item[\tt Set] $\mathcal{C'} \leftarrow \mathcal{C}$
			\item[\tt Update] \quad $\mathcal{C'}$ by removing each location in any partition with probability $par_c$.
			\item[\tt Set] semantic seed $semseed \leftarrow seed$
			\item[\tt Update] \quad $semseed$ by replacing all locations $r$ in the trace with the set of locations in their corresponding clusters, i.e., replace $r$ with $\mathcal{C'}_i$ where $r \in \mathcal{C'}_i$
			\item[\tt Update] \quad $semseed$ by removing the location $r = seed(t)$ at any time $t$ with probability $par_l$
			\item[\tt Update] \quad $semseed$ by merging locations that are located with time distance $\Delta t$ with probability ${par_m}^{\Delta t}$
			\item[\tt Set] $fake \leftarrow$ \textsf{HMMDecode}$(semseed, \langle \bar{p}, \bar{\pi} \rangle )$
			\item[\tt If] \textsf{PrivacyTest}$(fake, seed, par_s, par_i, par_d)$\\
				\texttt{Set} \quad $\mathcal{F} \leftarrow \mathcal{F} \cup fake$
		\end{description}
\end{description}
\texttt{Return} $\mathcal{F}$
}}}
}
\caption{Fake traces generation algorithm. We present it simplified for the case with a single time period.}
\label{fig:alg}
\end{figure}

In this section, we present the details of our algorithms for sampling fake traces. Figure~\ref{fig:alg} presents a high-level view. The process of generating and using fake traces are completely separate. When a set of fake traces are generated, they can be used in any protection mechanism accordingly.

\subsection{Transform Traces into Semantic Domain}
We assume that we have a dataset $\mathcal{S}$ of real traces that we use as seed to generate fake traces. Each seed trace in the dataset comes from a different individual. Generating a fake trace starts by transforming a real trace (taken as seed) to a semantic trace. To this end, we require to know the semantic coordinates of the seed trace. We compute the semantic similarity between all locations in $\mathcal{R}$, and create a location semantic graph $G\langle\mathcal{R},E,w\rangle$ such that the vertices are in $\mathcal{R}$ and the weight $w_G(r, r')$ on the edge between locations $r$ and $r'$ is the weighted sum of the number pairs of users $u$ and $v$ for whom $r$ and $r'$ is semantically mapped (i.e., $r = \sigma_u^v(r')$), weighted according to their similarity. 
 Then, we create the equivalent semantic classes $\mathcal{C}$ by running a clustering algorithm on this graph. For this purpose, we make use of k-means clustering algorithm, and we choose the number of clusters such that it optimizes the clustering objective. We present the sketch of this algorithm in Figure~\ref{fig:alg}-\textsf{SemanticClustering}$()$.

\begin{figure}[t]
\centering
\includegraphics[width=0.85\columnwidth]{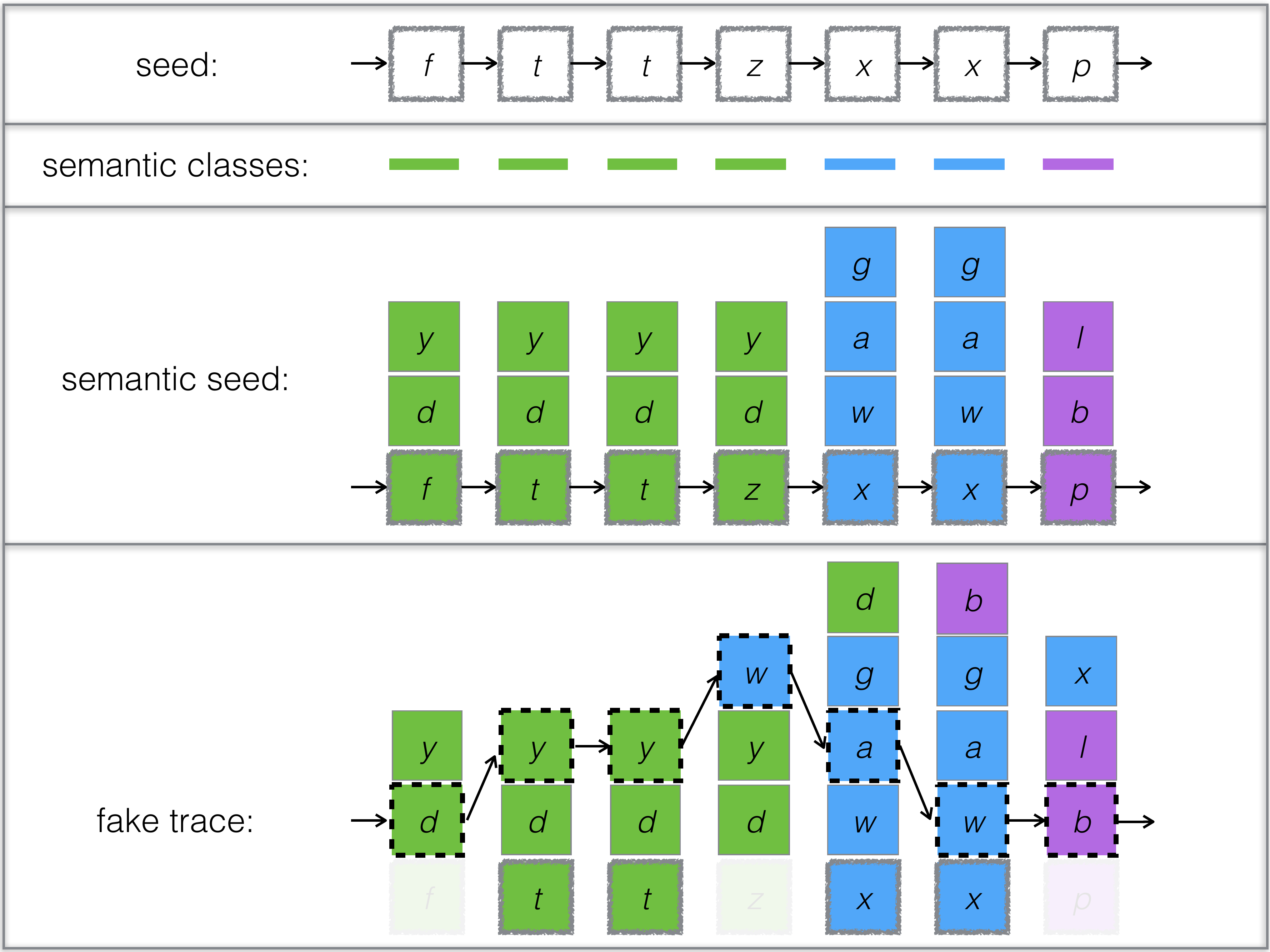}%
\caption{A sketch of generating a fake trace from a seed. Each location is represented by an English letter in a box. The semantic class associated with each location is represented by a different color. The semantic seed trace is a trace that includes the locations in the seed along with other locations in the same cluster at each time instant. Here, locations are clustered as $\{y, d, f, t, z\}, \{g, a, w, x\}, \{l, b, p\}$. To generate a fake trace, we first probabilistically remove the seed location and probabilistically merge subsequent classes. In our example, $f, z, p$ are removed, and $w, d, b, x$ are merged into their neighboring visited clusters. We then run a decoder to generate a probable trace given the possibility of choosing from all available locations at each time instant. The fake trace is shown with dashed boxes. A rejection test will run on this trace to guarantee its privacy compliance.}%
\label{fig:procedure}%
\end{figure}

We then convert the seed location trace $seed$ into its corresponding semantic trace $semseed$ by simply replacing each location in the trace with all its semantically equivalent locations (according to the semantic classes $\mathcal{C}$). Figure~\ref{fig:procedure} depicts an example of such a semantic seed. Intuitively, this composite trace encompasses all possible geographic traces that have a high semantic similarity to the original seed trace.
 To be more flexible with respect to the traces that we can generate, we add some randomness to the semantic seed trace. In the transformation process of the seed trace into the semantic trace, we sub-sample locations from the semantic classes as opposed to using them all. For privacy reasons, we remove each location in a cluster with probability $par_c$. The result is a new cluster $\mathcal{C'}$. We also allow locations of different classes to merge into each other closer to the time instants where the user moves from one class to the other. We implement this by merging a location between two cluster visited $\Delta t$ time instants away with a geometric probability ${par_m}^{\Delta t}$. 

\subsection{Sample a Trace from the Semantic Domain}
Any random walk on the semantic seed trace that crosses the available locations at each time instant is a valid location trace that is semantically similar to the seed trace. However, the synthetic traces we want to generate also need to be geographically consistent with the general mobility of people in the considered area.

We cast the problem of sampling such traces as a decoding problem in Hidden Markov Models (HMMs) \cite{rabiner1989tutorial}. The symbols are locations, the observables are the semantic classes (or the set of semantically equivalent locations in the same class), and the transition probability matrix is our aggregate mobility model. We construct the aggregate mobility model by averaging over the mobility models of all traces in dataset $\mathcal{S}$, as well as giving a small probability to the possible movements between locations according to their distance and connectivity. More precisely, we compute the aggregate transition probability $\bar{p}_r^{r'}$ as $\frac{\sum_{u\in\mathcal{S}} p_r^{r'}(u) + \epsilon \cdot {max(1, d(r, r'))}^{-2}}{z_r}$, where $\epsilon$ is a small constant, and $z_r$ is the normalizing factor. We compute the aggregate visiting probability $\bar{\pi}^r$ as the average of $\pi^r(u)$, for all $u\in\mathcal{S}$. The probability distribution $\bar{\pi}$ is also the steady state probability distribution of $\bar{p}$.

By decoding the semantic trace into geographic traces using HMMs, we generate traces that are probable according to aggregate mobility models, i.e., there could be one individual who prefers to take that trajectory. There are different HMM decoding algorithms. We make use of the Viterbi algorithm which is a dynamic programming algorithm to generate the most probable trace given the observation (i.e., the sematic seed trace) \cite{viterbi1967error}. More precisely, Viterbi finds $$\arg\max_{fake} Pr\{fake | semseed(t), \langle \bar{p}, \bar{\pi} \rangle \}$$ assuming that $fake(t)$ can only choose from locations in $semseed(t)$. Finding the most likely fake trace is equivalent to finding the shortest path in an edge-weighted directed graph where each location at time instant $t$ is linked to all locations at the subsequent time in the semantic seed trace.

By using this encoding technique, we make sure that the sampled trace is consistent with the generic mobility and has a significant probability of (geographically) being a real trace. However, Viterbi produces (only) one trace, hence we cannot directly generate multiple fake traces. To address this issue, we add randomness to the trace reconstruction of Viterbi. We modify the Viterbi algorithm, which originally, at each step (time instant) selects the most probable location in the path; we add some randomness to the probabilities such that the algorithm does not deterministically select the most probable location. More precisely, we slightly perturb the probabilities in such a way that Viterbi selects randomly among a set of locations that are close in probability to the most probable location. We implement this idea by choosing a parameter $par_v$ and multiplying all the probabilities of moving from one location to the next with a random number between $1$ and $par_v$. 

\subsection{Threat Model}

The threat against fake trace generating algorithms is twofold. (1) One threat is directly related to the adversary who wants to filter out traces that are fake and to find out the true location of mobile users ({\em localization} attack), e.g., when it is used in location based services to hide the true location of the user. For this attack, we assume the adversary has a background knowledge on general mobility models of users in the considered area. In Section~\ref{sec:evaluation_lbs}, we quantify the success rate of an adversary in localizing users while fake traces are used as a defense mechanism.  (2) The other threat is secondary as it is related to the algorithm that generates the fakes, and comes from the adversary whose goal is to identify the real individuals whose trace are used as seed ({\em membership inclusion} attack). In the next subsection, we enforce a privacy property that protects against the second threat. We guarantee plausible deniability for seed contributors independently from the adversary's knowledge.

\subsection{Privacy Tests}
\label{sec:privacy-test}
In the end, we want to make sure that the generated fake traces do not leak information about the seed trace. We design tests to protect against the following threat model. We assume the adversary has access to traces (or mobility patterns) of some individuals that might overlap with the set of individuals who contribute to the seed dataset. The attacks depend on the scenario in which a fake trace is used. The adversary might be interested in separating fake from real (in LBS scenario) or finding the seed from which a fake trace is generated (in trace publishing scenario). To protect against such threats, as the last step of our process, we run a \textsf{PrivacyTest}$()$ on each of the generated fake traces.

{\bf --} We compute the geographic similarity of each fake trace to the seed trace and reject the fake trace if its similarity is higher than a threshold $par_s$. This makes sure that the fake trace does not {\em statistically} leak information about the mobility of the individual behind the seed trace. We also reject a fake trace if its intersection with the seed trace is larger than $par_i$. This makes sure that the exact locations visited by the individual are not present in the fake trace. These tests provide the privacy guarantee with respect to information leakage of visited locations. 

{\bf --} We also use another notion of privacy, which is more of relevance in the case of publishing a location dataset. Specifically, we want to defend against {\em membership inclusion} attacks, in which an adversary wants to infer whether a particular individual's data was included in the seed dataset. Therefore, we design another privacy test that guarantees {\em plausible deniability} for those whose trace was used as seed. The main idea is that a fake trace should be as semantically similar to its own seed as to some other real traces which are not included in the seed dataset, so that an adversary cannot certainly infer that a particular individual was in the seed dataset and de-anonymize the seed contributor. Intuitively, if the semantic similarity of a fake trace to its own seed is comparable to its semantic similarity with some non-seed real trace, then the fake trace could have been generated from either of them. To enforce this property, we test if for a generated fake trace $f$ (that was generated from a seed trace $s$), there are some other real trace $s'$ such that their differential semantic similarity is bounded. $$|\mathsf{sim_S}(s, f) - \mathsf{sim_S}(s', f)| < par_d$$

We can enforce this to hold for a minimum $k-1$ number of alternative $s'$ traces (from which we are not going to publish fake traces). Thus, there is at least one real trace outside the set of seeds associated with fake traces that could have produced each releasable fake trace. More generally, we enforce the size of anonymity set to be $k$. This property, which provides {\em plausible deniability}, is conceptually related to, but weaker than, Differential Privacy (DP)~\cite{Dwork06}. Indeed, this is one kind of guarantee that differential privacy is meant to provide.\footnote{\scriptsize DP can be thought of as providing plausible deniability by thinking of the differing elements between the two neighboring datasets as two possible inputs of the same user. When DP is satisfied, the output distribution is approximately the same, so that that user can plausibly pretend having used any one of the two inputs.} That said, we enforce this property for {\em actual} seeds in the datasets. Thus, by looking at a fake trace the adversary cannot surely conclude that a particular trace was in the seed dataset, because there exist at least $k-1$ other traces that could have seeded the same fake trace.

Each time we generate a new fake trace that passes the privacy tests, we compute its likelihood based on the aggregate mobility model. One can then randomly sample from the bag of fake traces based on this likelihood. The traces that are generated according to this process do not leak information about the seed traces, yet they share their average geographic features and semantic features.

\subsection{Discussion}\label{sec:alg-disc}
The fake generation process, which results in a pool of fake traces having passed the privacy test, is run offline on powerful machines, before the users' device retrieve and use such fakes.  Therefore, this computational burden is not placed on the users' device. Nevertheless, the generation process is reasonably efficient: the computation of both the aggregate mobility and the semantic clustering needs to be done only once for each input set of real traces. The former takes time $O(S L + (RT)^2)$ where $S=|\mathcal{S}|$ is the number of seed traces, $L$ is the length (i.e., number of events) of each seed trace. 
The latter is dominated by $S(S-1)$ semantic similarity computations (e.g., each taking $O(T R^3)$ in the zeroth-order case) and one clustering operation. Excluding the final clustering, this step is embarrassingly parallel: the semantic similarity for any two users $u$, $v$ can be computed independently. Also, if a few input traces are added, both the aggregate statistics and the semantic clustering can be updated and do not need to be recomputed from scratch.  
Once the semantic clustering has been computed, an arbitrarily large number of fakes for each seed can be generated. This process is also embarrassingly parallel, since each fake can be generated independently of other fakes for that seed, and other seeds. 

The algorithm of Figure~\ref{fig:alg} works if the input dataset contains at least two seed traces. However, its quality will be high if, among the seed traces: the coverage of the location set $\mathcal{R}$ is high (not necessarily complete); the semantic similarity is high; and the geographic similarity is low.

\section{Evaluation}\label{sec:evaluation}
\begin{figure}[t]
	\hfill\subcaptionbox{{\scriptsize Visited Locations. The size of locations are proportional to their total visits.} \label{fig:visited-locs}}	
	{\includegraphics[width=0.475\linewidth]{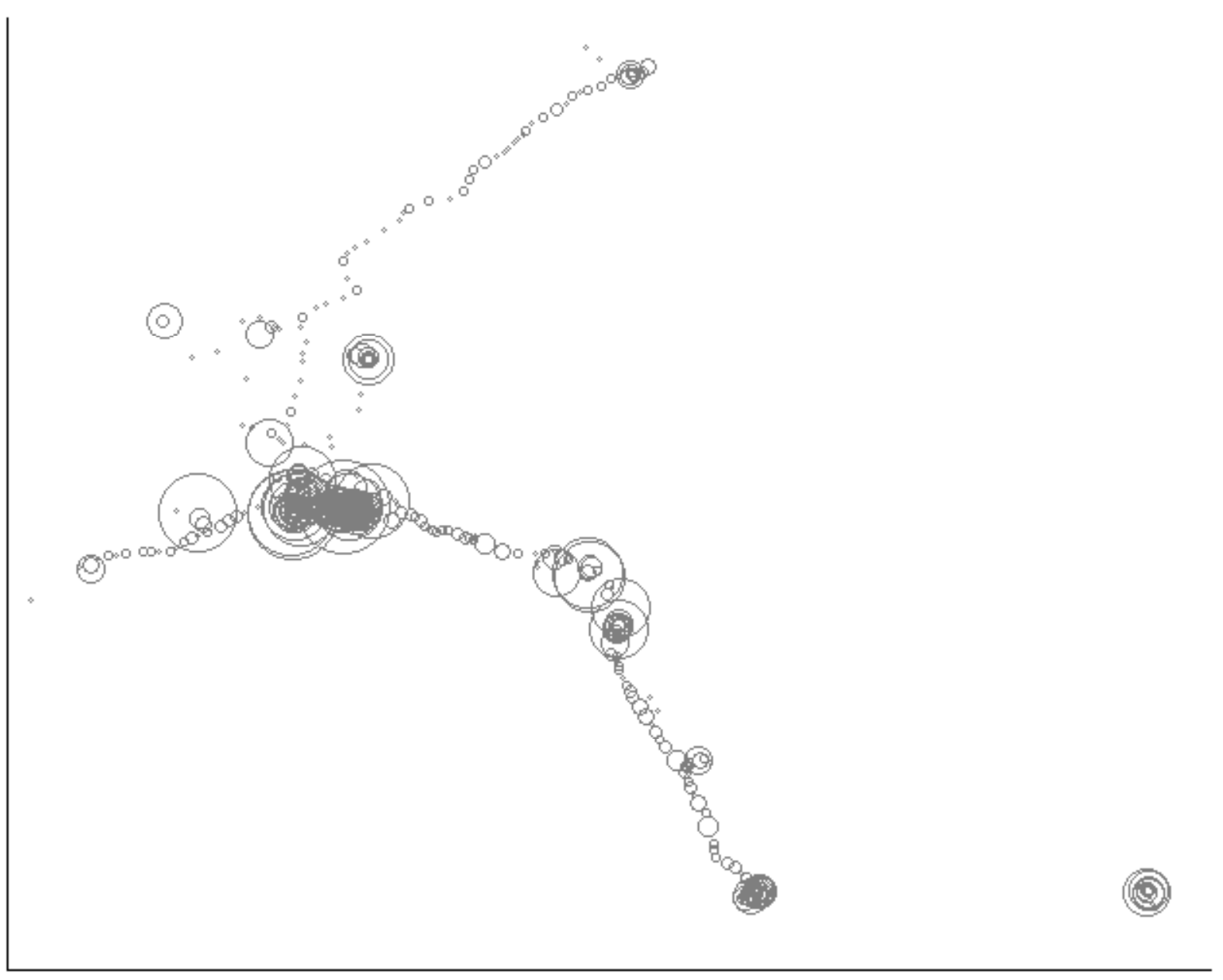} }%
	\hfill\subcaptionbox{\scriptsize Visited locations colored according to their semantic clustering (20 clusters). \label{fig:semantic-clusters}}
	{\includegraphics[width=0.475\linewidth]{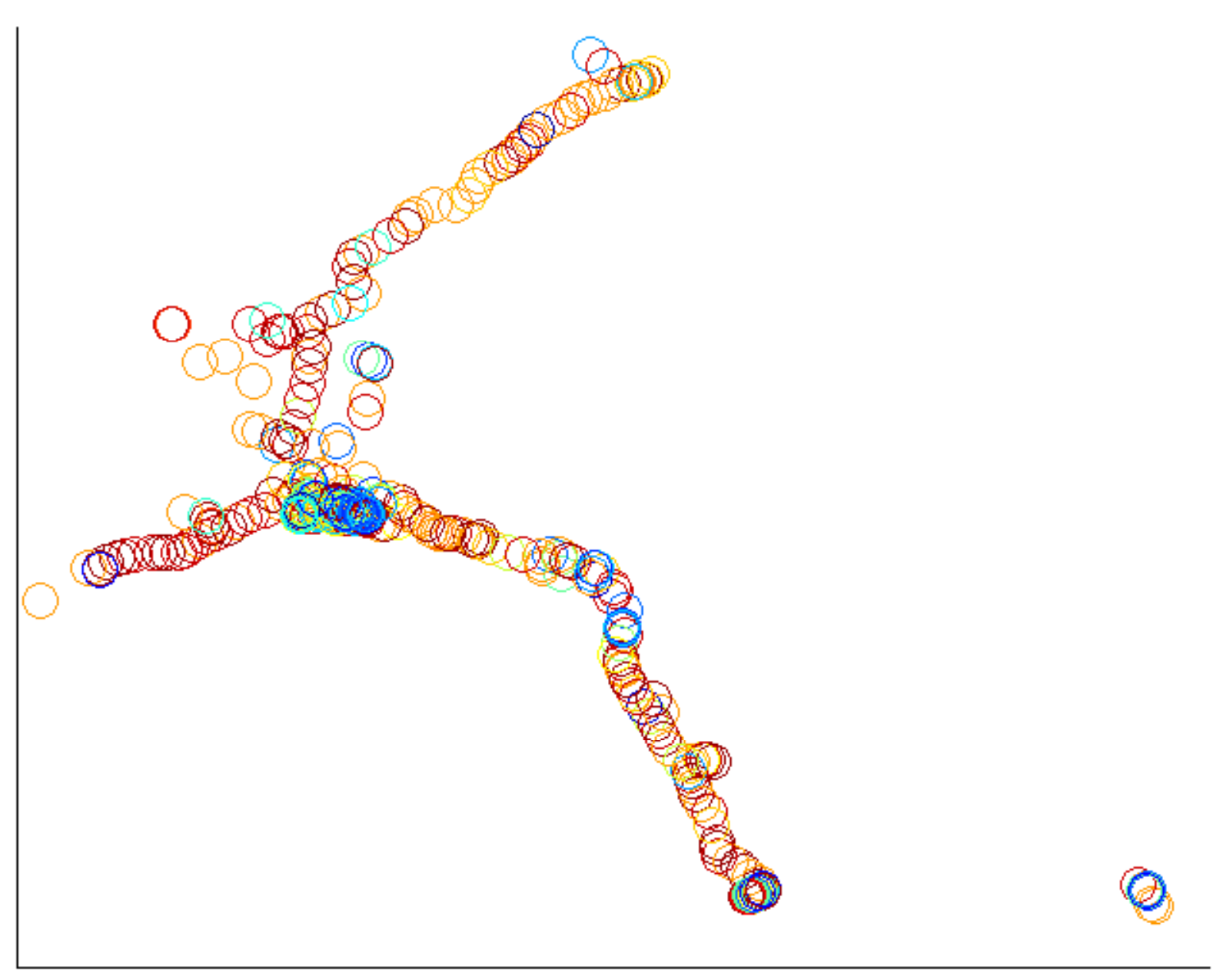} }%
	\caption{400 locations visited around Lausanne and nearby towns by the 30 users. Some users commute between two towns whereas the majority of them live and work in the same city of Lausanne (the area with higher concentration). \label{fig:dataset}}
\end{figure}

In this section, we run our algorithms on a set of real location traces and evaluate their utility and privacy in two scenarios: publishing a location dataset, and sharing locations with a location-based service.

\subsection{Dataset}
\label{sec:dataset}
The dataset we use for the evaluation is collected through the Nokia Lausanne Data Collection Campaign\footnote{\scriptsize http://research.nokia.com/page/11367} (see~\cite{kiukkonen2010towards}). We prepare the dataset for our needs in two phases, filling gaps in the traces and discretizing the time and location.

The raw dataset contains events of three types: GPS (the GPS position of the user is known), WLAN (the SSID and signal strength of a set wireless networks which surround the user are known), and GSM (the identifier of the GSM base station to which the user's phone is associated is known). In the first phase, we compute valid traces (out of possibly partial traces) by aggregating events and filling gaps. We do this by interpolating along the path of consecutive GPS points and using the WLAN and GSM information.

In the second phase, we extract two days of traces for 30 users, such that each trace contains a sequence of 72 locations (i.e., one location is reported every 20 minutes). Some locations are visited very rarely only by very few users. Thus, we reduce the number of locations from 1491 to 400 by clustering close-by locations together. We use a hierarchical clustering algorithm for this purpose, in which the distance between two locations is taken to be proportional to both the Euclidean distance between the locations and the product of their weights (defined as the number of total visits to each location, for all users). This means that locations clustered together will tend be both geographical close and have few visits. The geographical distribution of visits of all users over the locations in the considered area is shown in Figure~\ref{fig:dataset}(a).

We computed the mobility profiles of all 30 users, and then the semantic location graph by calculating a similarity score for each pair of locations, averaged across all users. After clustering this semantic location graph, we obtained 20 location clusters. We choose this number of clusters as it provides optimal clustering i.e., it maximizes the ratio of inter-cluster similarity over intra-cluster similarity. This clustering is illustrated in Figure~\ref{fig:dataset}(b), where each location is drawn with the color of the cluster it belongs to. The figure allows us to distinguish some patterns, for example locations at the center of cities are mostly in blue, while many locations representing roads and highways are colored in red. Also notice that the semantic clustering does not seem to depend on the geographical distance of locations.

\begin{figure}[t]
	\hfill\subcaptionbox{\label{fig:real-geographic-similarity}}	
	{\includegraphics[width=0.48\columnwidth]{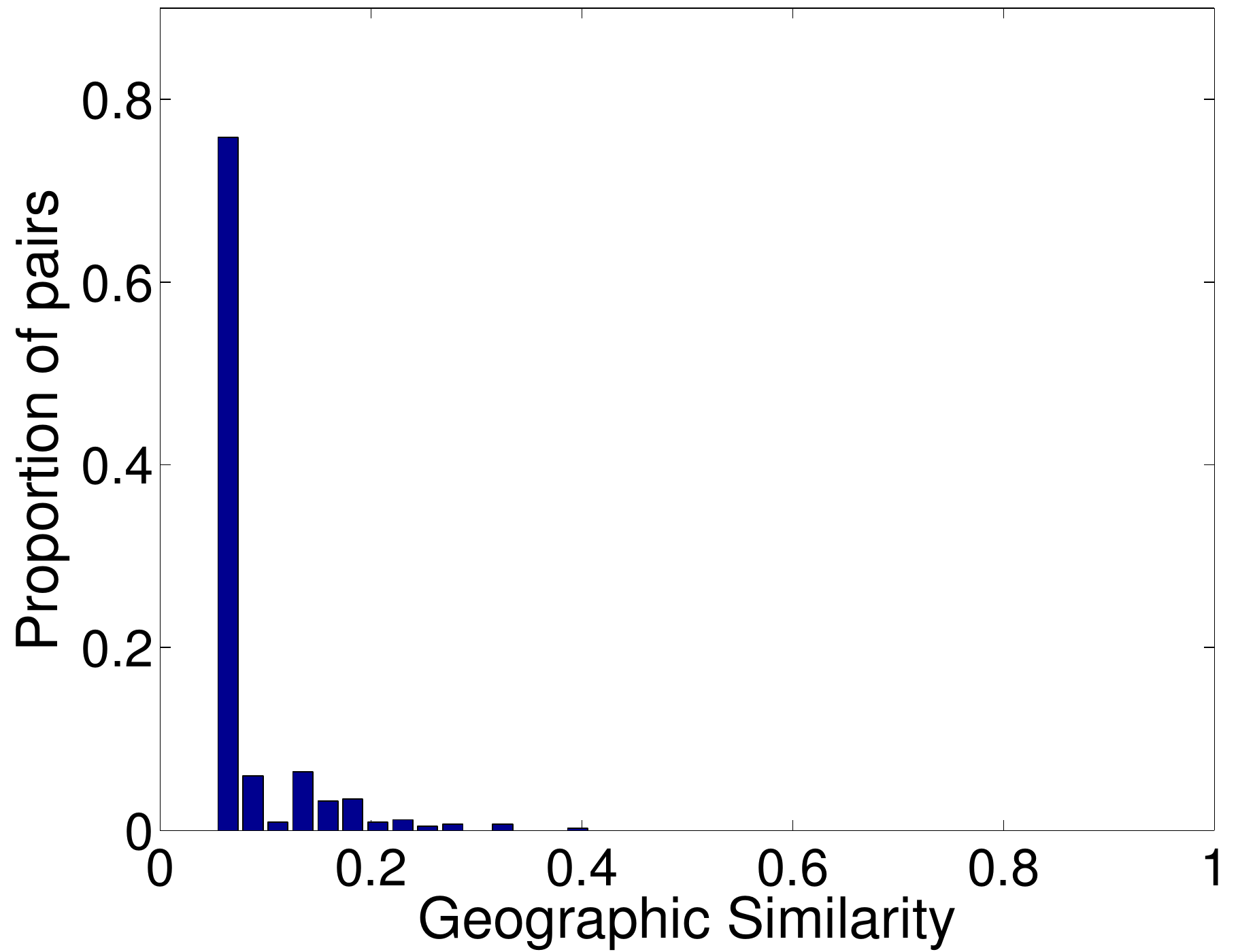}}%
	\hfill\subcaptionbox{\label{fig:real-semantic-similarity}}
	{\includegraphics[width=0.48\columnwidth]{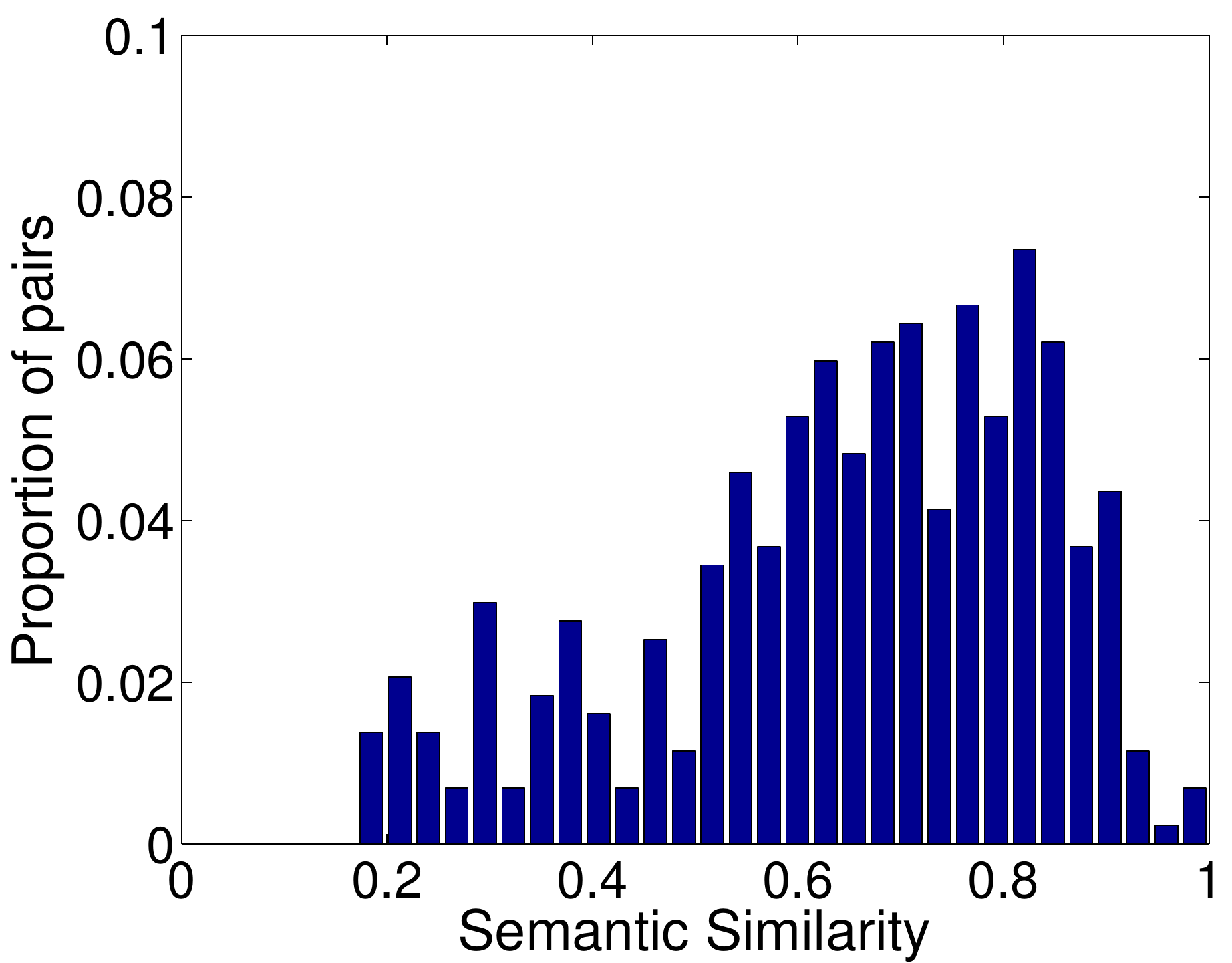} }%
	\caption{Normalized histogram of the (a) geographic similarity and (b) semantic similarity of all distinct pairs of 30 users. (a) Mobility models of different individuals is geographically very specific to themselves, i.e., they are unique. This is well reflected in the skewed distribution of geographic similarity towards very small values. (b) As hypothesized in this paper, majority of individuals have high semantic similarities with respect to their mobility models.\label{fig:real-gs-similarity}}
\end{figure}

To illustrate the difference between geographic and semantic similarities, we can compute those metrics pairwise over all 30 users.\footnote{\scriptsize Since both the geometric similarity, and the semantic similarity of a user with herself is 1.0, we exclude such pairs.} The result is shown in Figures~\ref{fig:real-gs-similarity}(a), and~\ref{fig:real-gs-similarity}(b). The first histogram shows that the 30 users are not strongly geographically similar to each other, except for a few pairs of users. This is expected given the range of locations they explore overall, as seen in Figure~\ref{fig:dataset}(a). On the other hand, the distribution of the semantic similarity across all distinct pairs of users has a larger variance; while some pairs of users are not similar at all (e.g., those with semantic similarity score of 0.2), a large number of users are highly similar.

\subsection{Fake Trace Generator Tool}
We build our tool to generate fake traces on top of the open-source Location Privacy Meter (LPM)~\cite{ShokriTLH11}. To exploit LPM's modularity we split our algorithm into modules. To implement the time-dependent sub-sampling of clusters and merging around transitions, and the transformation of users' actual traces into semantic traces, we use the location obfuscation mechanism feature of the tool. The reconstruction of geographically valid synthetic traces from the semantic traces is done using the Viterbi algorithm (implemented in LPM as a tracking attack). To cluster the semantic location graph, we employ the CLUTO toolkit \cite{ryen14multiusersearch}.

\subsection{Simulation Setup}
\label{sec:sim-setup}
As for the parameters of the \textsf{GenerateFake}$()$ algorithm, we set the location-removal probability $par_c$ to $0.25$, and we set the location merging probability $par_m$ to $0.75$. We set the probability $par_l$ of removing the true location visited in the seed to $1.0$.
We set the randomization multiplication factor for Viterbi randomization to 4. So, for each probability assigned to each location at each time instant, we multiply it with a randomly chosen number between 1 to 4. We set very tight values for the privacy parameters. We set $par_i$, the maximum intersection between fake and seed, to $0$. So, we do not tolerate any intersection between fake and seed. We set the geographic similarity threshold $par_s$ to $0.1$, and the differential semantic similarity threshold also to $0.1$.

For each of the 30 users, we generated about 500 fake traces. Out of those we randomly pick 50 traces (for each user) to be used for the datasets publishing scenario. For the LBS scenario, we sampled traces (for each user) according to the traces likelihoods, out of the pool of traces (for that user) which passed the privacy test.

Out of the two days of traces (each 72 timestamps, for each of the 30 users), we use the first day as the training dataset, and the second day as the testing dataset. We calculate the aggregate statistics and mobility profile of users on the training dataset, while we use the testing dataset to evaluate both the data publishing scenario and the attack for the LBS scenario. Unless otherwise stated, for all experiments, we consider a single time-period, and compute the zeroth order versions of the geographic and semantic similarities.

\subsection{Evaluation Metrics}
\label{sec:eval-metrics}
In the following two subsections we evaluate the use of fake traces in two popular scenarios: publishing a dataset of fake traces, and using fake locations along with real locations when accessing location-based services. In both scenarios, we evaluate our fake traces with respect to two metrics: {\em privacy} and {\em utility}. The privacy guarantee that we provide using our privacy rejection test applies to both scenarios. However, there are some differences in terms of the adversary model between different scenarios. There are therefore some additional considerations regarding the privacy of users in location-based services, e.g., their privacy against inference attacks, that we discuss in its corresponding subsection. The utility metric is very dependent on the application (scenario), hence is measured differently in each case. 
\begin{figure*}
	\centering
	\hfill
  \begin{minipage}[t]{0.3\textwidth}
    \includegraphics[width=\columnwidth]{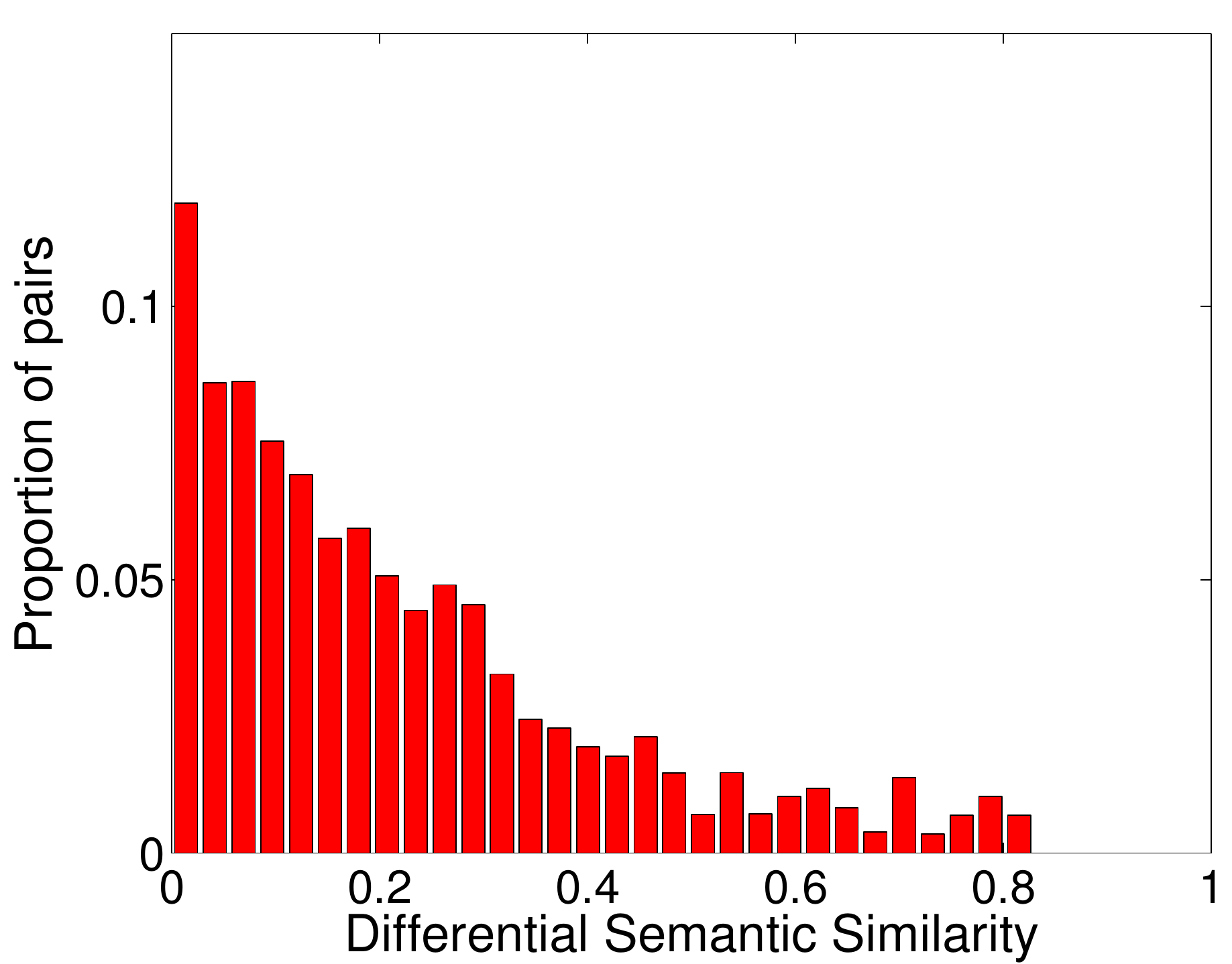}%
    \caption{Histogram of the differential semantic similarity between fake and real traces.     It presents the distribution of the absolute difference $|\mathsf{sim_S}(s, f) - \mathsf{sim_S}(s', f)|$, for all pairs of $f$ (plus its seed $s$) and $s'$.
    }
    \label{fig:deltas-fakeset2}%
  \end{minipage}\hfill
  \begin{minipage}[t]{0.3\textwidth}
    \includegraphics[width=\columnwidth]{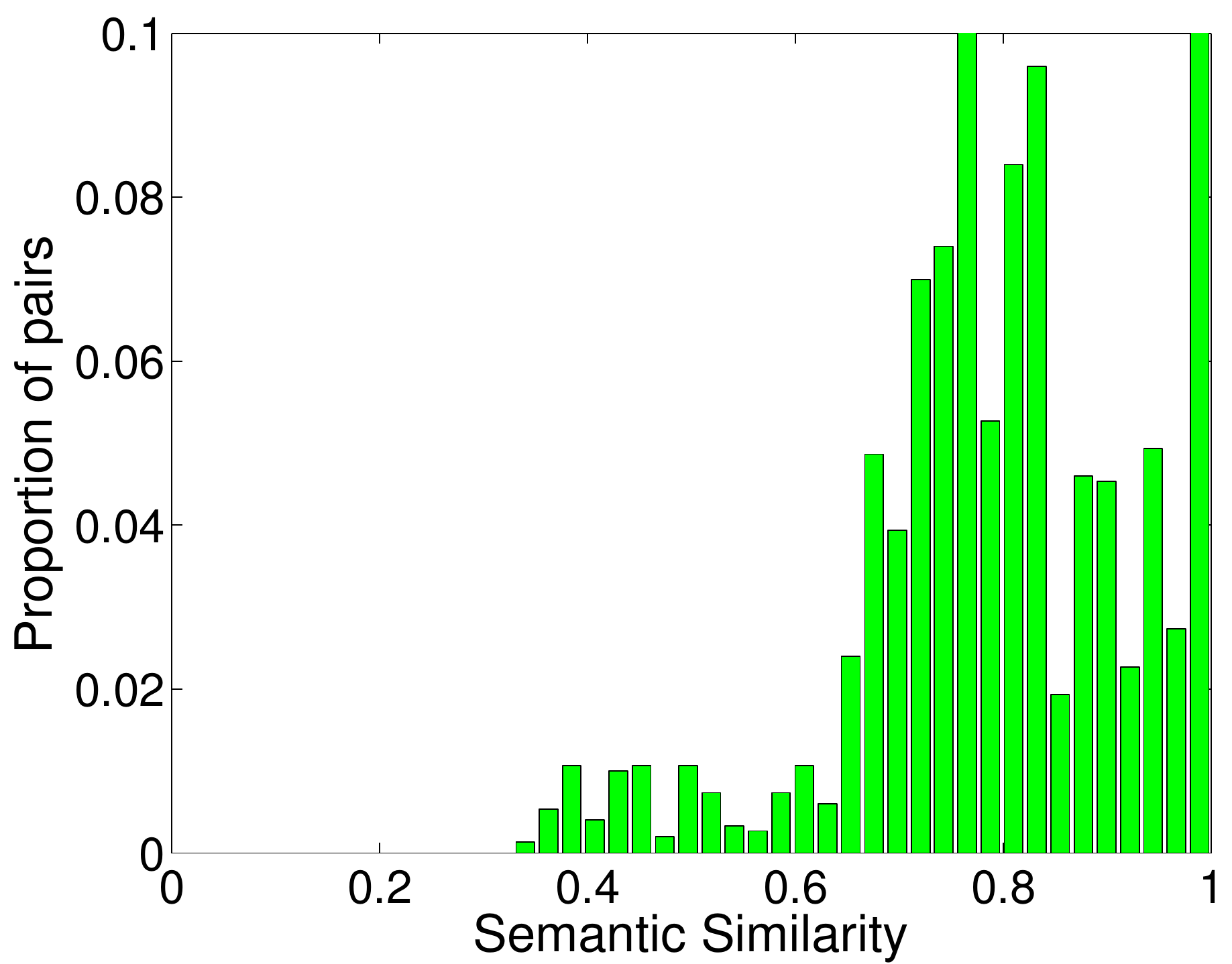}%
    \caption{Normalized histogram of the semantic similarity of all distinct pairs of: each of the 30 real traces, along with their associated fake traces.}
    \label{fig:real-fake-semantic-similarity}%
  \end{minipage}\hfill
  \begin{minipage}[t]{0.3\textwidth}
	\includegraphics[width=\columnwidth]{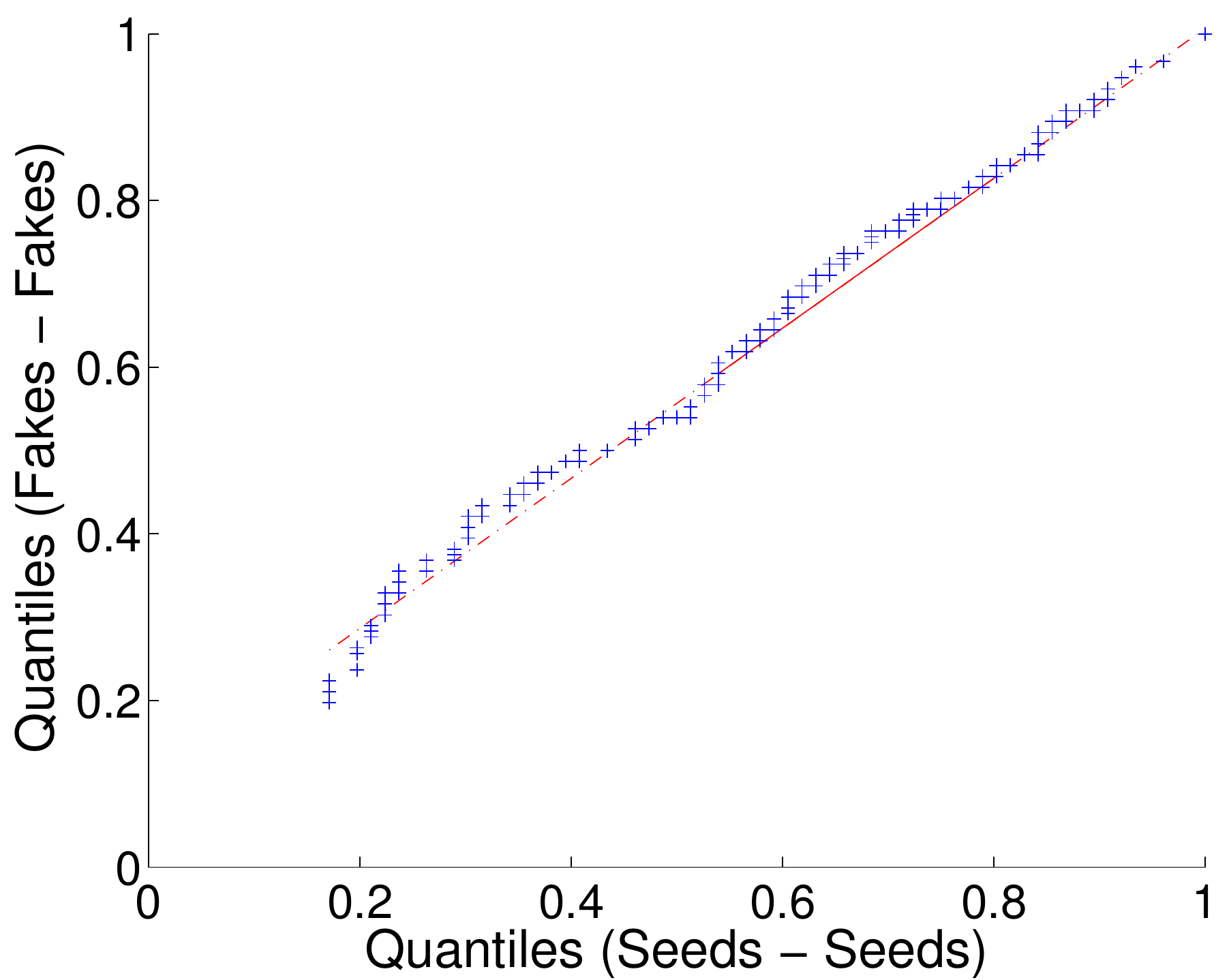}%
    \caption{Q-Q plot for comparing two distributions: semantic similarity among all real seed traces, and semantic similarity among all fake traces. The plot shows a very strong correlation between two distributions. 
    }
    \label{fig:fake-fake-semantic-similarity-qq}%
  \end{minipage}
\end{figure*}

\subsection{Scenario: Publishing Location Datasets}
\subsubsection{Setup} 
In this scenario, we assume that we generate a fake trace for some seed traces and publish them all in a dataset. We use some real traces in our dataset that we use as alternative seeds in the differential semantic similarity test. 

\subsubsection{Privacy} 
We assume an adversary wants to infer information about the true traces that have been used to generate the fakes. Due to the privacy test, location privacy of individuals who contributed to the seed dataset is guaranteed as fake traces do not intersect with the locations that are visited and does not even leak statistically about what could have been visited by those individuals. Moreover, due to the differential semantic similarity test, the seed traces are not the only traces that could have generated the released fake traces. 

Out of all fake traces that we generated from our 30 seed traces, on average $80\%$ of them could pass the geographic and intersection privacy tests with tight constraints ($par_i = 0$, i.e., no intersection allowed, and $par_s = 0.1$), so it is not difficult to generate fake traces that satisfy such privacy guarantees. Regarding the differential semantic similarity test, we should be able to find enough number of real traces as alternative traces that could have been the seed for releasing fake traces. In Figure~\ref{fig:deltas-fakeset2}, we show the difference between semantic similarity of a fake trace and its seed with the semantic similarity of the same fake trace and any other real trace in our dataset. The histogram shows that the majority of fake traces have very low similarity to real traces other than their seeds. 
This is due to the high semantic similarity between real traces (Figure~\ref{fig:real-semantic-similarity}) so it is not difficult to find potential alternative seeds for a fake trace. We set $par_d$ to  $0.1$ to obtain a high level of differential semantic privacy.

\subsubsection{Utility} 
To preserve the utility of the original traces, fake traces should share similar statistical properties with the real trace dataset. Note, however, that we would not expect all useful statistics to be preserved since some may be counter to our goal of preserving privacy. That is, certain geographic features are expected not to be preserved, due to the nature of the generation and the privacy test. For example, if a location is primarily visited by a single user in the real dataset, it is unlikely that the location would be visited with similar frequency by a (fake) user in the fake trace dataset. This is because if such a fake trace were generated from that seed, the privacy test would reject it. Nevertheless, we can evaluate to what extent certain useful statistics are preserved. 

To start, we compare the basic mobility statistics obtained from the real and fake datasets. We compute the aggregate mobility model for each fake dataset (we generate 10 of size 30), and compare its geographic similarity with the real dataset. More precisely, for a fake dataset $\mathcal{F}$, we compute $\langle \bar{p}_{\mathcal{F}}, \bar{\pi}_{\mathcal{F}} \rangle$ and compute its similarity to $\langle \bar{p}, \bar{\pi} \rangle$. The statistical similarity of $\bar{p}_{\mathcal{F}}$ with $\bar{p}$ over all fake datasets is $$[0.8061,\quad 0.8073,\quad 0.0060]$$ on (average, median, standard deviation), and the results for the statistical similarity of $\bar{\pi}_{\mathcal{F}}$ with $\bar{\pi}$ is $$[0.7856,\quad    0.7867,\quad    0.0152].$$ Both these results show a strong correlation between average/aggregate mobility information of real and fake datasets.

We then compare the location visiting probabilities of the real dataset and fake datasets. Namely, for each dataset we compute the spatial allocation, i.e., for each location (from least to most popular, for that dataset), we calculate the number of visits spent in that location across all traces in the dataset. We then normalize this quantity to obtain a probability distribution over locations (sorted by popularity), i.e., for each location we have the probability of a random visit to that location. From these distributions, we compute the KL-divergence of the real (training) dataset to each of our fake datasets, and to a variety of baselines.\footnote{\scriptsize Because the KL-divergence is only defined at points where the second distribution is not zero, unless the first is also zero, we set all zero probabilities to $\epsilon = 0.1$, before normalizing. This is required because there may be locations which are visited in the fake dataset but not in the real dataset, or vice-versa.}  The results are shown in Table~\ref{tbl:vistprob-kl}. Since the KL-divergence is not upper bounded, we use as baselines the KL-divergences of the real (training) dataset to the following distributions: real testing dataset; uniform visiting probabilities; and single location visiting. We see that while the KL-divergence of the real (training) dataset to the real testing dataset is smaller than that to the fake datasets, the latter is also significantly smaller than both the the KL-divergences to the uniform visiting baseline and the single location visiting baseline. This indicate that a lot of information is preserved in the fake datasets. Next, we repeat the previous calculation of KL-divergence, but considering only visits to the $50$ most popular locations (of each dataset). Table~\ref{tbl:vistprob-kl-50mp} shows the results: the information is almost as well preserved in the fake datasets than compared to the real testing dataset.

We also compare the users time allocation of the real and fake datasets. Namely, for each dataset and each user, we calculate the time spent at each location, among the locations visited. That is, we calculate, for the three most popular locations of that user, what proportion of the time is spent in each. 
We perform this calculation across all 30 users and normalize the result. We compare this distribution for the real and fake datasets.
Table~\ref{tbl:timealloc-kl} shows the KL-divergence of the real (training) dataset to the fake datasets and baselines: real testing dataset; uniform time allocation (each user spends $1/k$ proportion of time at each of the $k$ locations); random time allocation (each user spends a uniformly random proportion of time at the location). 
This statistic is highly preserved in the fakes; sometimes the fake datasets' distribution is closer to that of the real (training) dataset, than the distribution of the real testing dataset is.

The previous results provide confidence that useful information is indeed preserved in the fake traces dataset. That said, our original goal was to preserve utility in the sense of semantic similarity, so it sensible to wonder how close we are to that goal. To determine this, we first compute the semantic similarity of each fake trace with its own seed trace to check if the semantic features of the original traces are indeed preserved. Figure~\ref{fig:real-fake-semantic-similarity} illustrates the distribution of this value over all fake traces. Clearly, the distribution is biased towards higher similarity values. So, the fake traces considerably preserve the semantic features of the real traces.

Another type of statistics that we would expect the set of fake traces to preserve is the inner similarity between the set of traces. In Figure~\ref{fig:fake-fake-semantic-similarity-qq}, we present the correlation between two distributions: semantic similarity among real traces, and semantic similarity among fake traces. The Q-Q plot shows a significant correlation between these two distributions; they are strongly linearly related. This reflect that in addition to maintaining the information about each seed, we also preserve the statistical relation among the traces.

Results show that fake traces cannot be distinguished from the real ones if it appears among some real traces. This is because the relation between a fake trace and the distribution of real traces is largely indifferent from that of a real trace with respect to both semantic and geographic features. 
\begin{table}[b!]
\centering
    \begin{tabular}{|c |*{5}{c|}} 
    	\hline
    	Testing & \multicolumn{2}{|c|}{Fakes} & Uniform & Single \\ \hline
		\multirow{2}{*}{0.0377} & Mean & Std & \multirow{2}{*}{1.1918} & \multirow{2}{*}{4.6666} \\  \cline{2-3}
		& 0.3841 & 0.0432 &  &  \\ \hline
	\end{tabular}
\caption{KL-divergence of the location visiting probabilities of the real (training) datasets against the 10 fake datasets, and various baselines. ``Testing'' is the testing portion of the real dataset (see Section~\ref{sec:sim-setup}); ``Uniform'' is the uniform distribution over all locations; and ``Single'' is the distribution where all users always visit the same location.}
\label{tbl:vistprob-kl}
\end{table}
\begin{table}[b!]
\centering
    \begin{tabular}{|c |*{5}{c|}} 
    	\hline
    	Testing & \multicolumn{2}{|c|}{Fakes} & Uniform & Single \\ \hline
		\multirow{2}{*}{0.0215} & Mean & Std & \multirow{2}{*}{0.2040} & \multirow{2}{*}{5.1131} \\  \cline{2-3}
		&  0.0289 & 0.0086 &  &  \\ \hline
	\end{tabular}
\caption{KL-divergence of the 50 most popular location visiting probabilities of the real (training) datasets against the 10 fake datasets, and various baselines (see Table~\ref{tbl:vistprob-kl}).}
\label{tbl:vistprob-kl-50mp}
\end{table}
\begin{table}[b!]
\centering
    \begin{tabular}{c |*{6}{c|}} 
    	\cline{2-6}
    	& Testing & \multicolumn{2}{|c|}{Fakes} & Uniform & Random \\ \hline
		\parbox[t]{2mm}{\multirow{2}{*}{\rotatebox[origin=c]{-90}{1st}}} & \multirow{2}{*}{0.0189} & Mean & Std & \multirow{2}{*}{0.1652} & \multirow{2}{*}{0.6794} \\  \cline{3-4}
		&  &  0.0125 & 0.0022 &  &  \\ \hline
		\parbox[t]{2mm}{\multirow{2}{*}{\rotatebox[origin=c]{-90}{2nd}}} & \multirow{2}{*}{0.0026} & Mean & Std & \multirow{2}{*}{0.0778} & \multirow{2}{*}{0.5360} \\  \cline{3-4}
		&  &  0.0092 & 0.0031 &  &  \\ \hline
				\parbox[t]{2mm}{\multirow{2}{*}{\rotatebox[origin=c]{-90}{3rd}}} & \multirow{2}{*}{0.0114} & Mean & Std & \multirow{2}{*}{0.0779} & \multirow{2}{*}{0.5092} \\  \cline{3-4}
		&  &  0.0089 & 0.0036 &  &  \\ \hline
	\end{tabular}
\caption{KL-divergence of the the users time allocation distribution among the three most popular locations (of each user) of the real (training) datasets against the 10 fake datasets, and various baselines. }
\label{tbl:timealloc-kl}
\end{table}
\subsection{Scenario: Using Location-based Services}\label{sec:evaluation_lbs}
The utility and privacy evaluation for the publishing dataset scenario applies to the case where traces are shared with a service provider. However, we can perform a more specific analysis on the fake locations when they are shared in a new setting. We present the details of how fake locations are used to protect location privacy, and how they perform against inference attacks despite the fact that they have passed privacy guarantee tests.

\subsubsection{Setup} 
We assume a user shares her current location with a location-based service with a probability $\beta$ (set to $0.5$ in our case). The service provider, in return, provides the user with contextual information about the shared locations (e.g., list of nearby restaurants, current traffic information on the road). The service provider would receive a sequence of locations that are visited by the user at different time instants. To protect her location privacy, i.e., hiding her location at the time of access to the LBS and also preventing the inference of the full trajectory, the user sends a number of fake locations along with her true location. We assume the user has access to some full fake traces, and at any time instant $t$, when she is accessing the server, she consistently adds the locations visited at $t$ on each fake trace to her actual location at $t$ and sends them to the server. The service provider responds to each of these location queries, and the user needs to filter out the results associated with fake locations to obtain the information about her true location.

We evaluate a few other methods to generate fake locations along with our method for comparison.

\begin{itemize}[noitemsep,nolistsep]
	\item Uniform IID: We sample each fake location independently and identically distributed from the uniform probability distribution. 
	\item Aggregate Mobility IID: We sample each fake location independently and identically distributed from the aggregate mobility probability distribution $\bar{\pi}$.
	\item Random Walk on Aggregate Mobility: We sample a fake trace by doing a random walk on the set of locations following the probability distribution $\bar{p}$.
	\item Random Walk on User's Mobility: We do a random walk on the set of locations following the probability distribution $p(u)$ to generate a fake trace.
\end{itemize}

\subsubsection{Privacy} 
We assume the adversary wants to filter out the fake locations and to find the true sequence of locations that are visited by the user. The privacy metric that we use is based on the error of adversary in his inference attack \cite{ShokriTLH11}. Put simply, the fraction of true locations that are missed by the adversary is our privacy metric. More precisely, the metric is the probability of error of inference attack on guessing the correct location. We assume the adversary makes use of the aggregate mobility model $\langle \bar{p}, \bar{\pi} \rangle$ to single out the true locations and reconstruct the true location of the user.

\subsubsection{Utility} 
There is an overhead to these privacy-preserving mechanisms, as a user has to send more than one query to the server to get the results for one query. This can be interpreted as utility loss, and so we define two metrics for (the lack of) utility. The first is the number of distinct locations sent by the user at each time (diversity overhead). Note that this number can be less than the number of fake traces as they might intersect at the times of connection to the server. Additionally, some service providers (e.g., Google Now) might profile the user over time based on the type of locations she visits, in order to provide recommendations or reminders. In these cases, the queries that are sent to the server can pollute the profile of the user hence reduce the predictability power of the service provider. For this, we use the number of (distinct) semantic clusters among the locations sent by the user at each time (semantic overhead). 
\begin{figure}[t]
	\hfill\subcaptionbox{\label{fig:locationprivacy1}}	
	{\includegraphics[width=0.48\columnwidth]{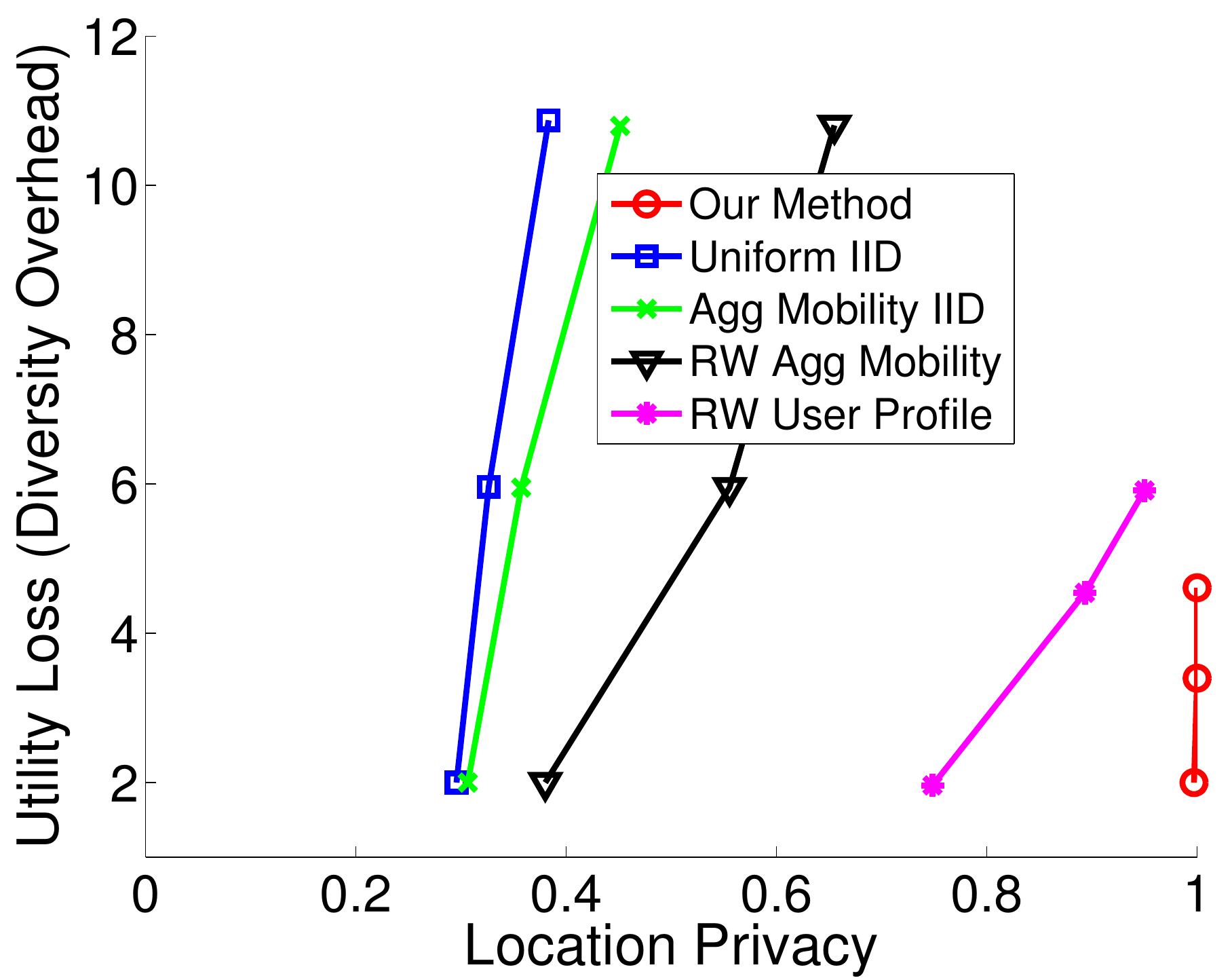}}%
	\hfill\subcaptionbox{\label{fig:locationprivacy2}}
	{\includegraphics[width=0.48\columnwidth]{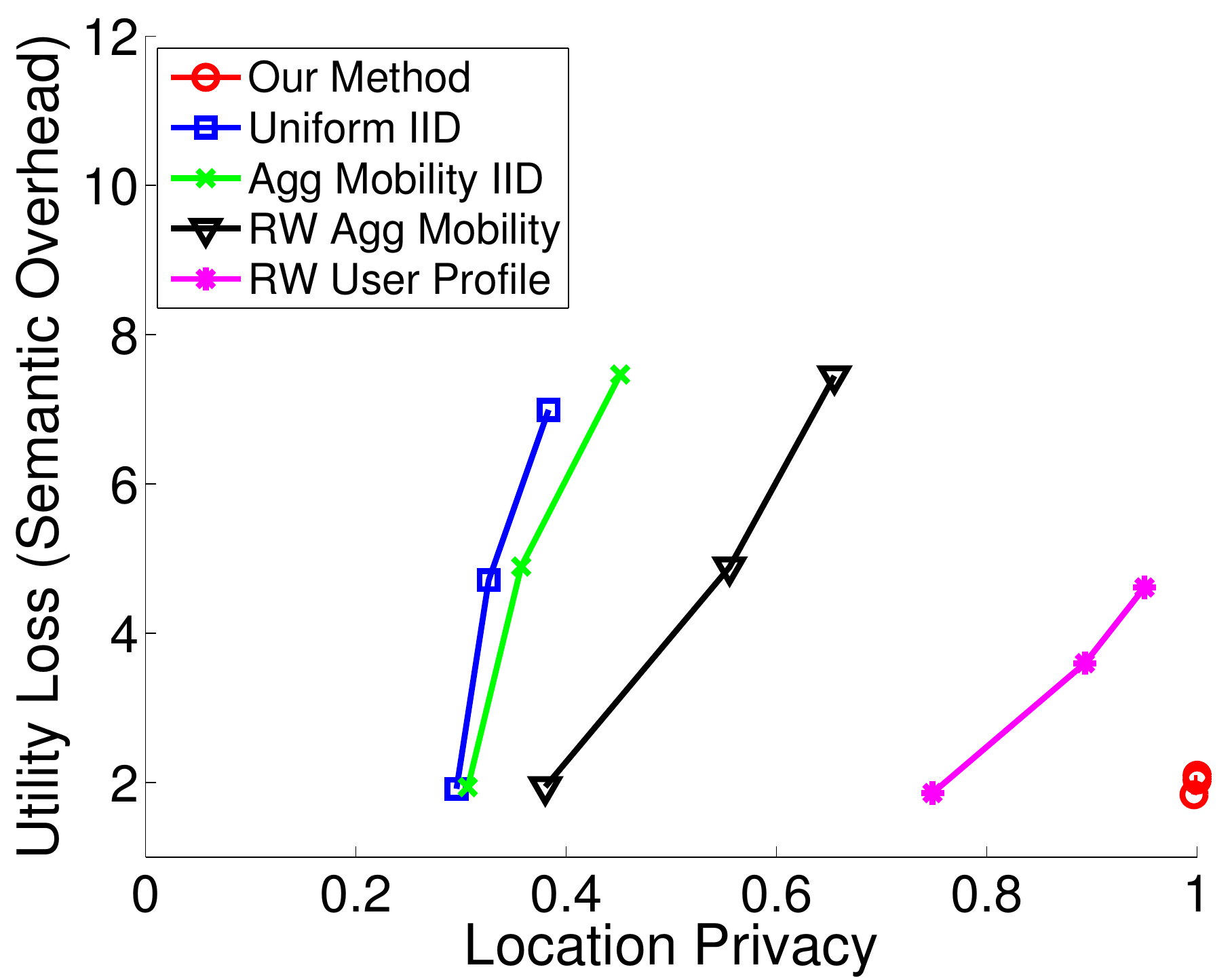} }%
	\caption{Location privacy versus utility loss for different fake generating algorithms. The privacy is measured as probability of error of adversary in guessing the correct location of users. We plot the median location privacy across all 30 users. The probability of connecting to the LBS is set to $0.5$, so one half of the time instants the users connect to the server. We evaluate the use of $1, 5, 10$ fake traces, hence three dots for each algorithm. (We repeated the experiment 20 times and took the average: 4 times with a different selection of fake traces, and for each of such selection, 5 times to eliminate the randomness of exposures.) The utility loss is the number of distinct locations that are sent to the server (\ref{fig:locationprivacy1}), and the number of (distinct) semantic classes (i.e., clusters) exposed for each event (\ref{fig:locationprivacy2}). 
	 \label{fig:locationprivacy}}
\end{figure}

\subsubsection{Results}
Figure~\ref{fig:locationprivacy} shows the tradeoff between location privacy and utility for various methods of generating fake traces. We evaluate the utility loss in terms of two metrics: diversity overhead (Figure~\ref{fig:locationprivacy1}), and semantic overhead (Figure~\ref{fig:locationprivacy2}). We evaluate the privacy for exposure probability $\beta = 0.5$, and three different number of fake traces: $1, 5, 10$. Although the number of fake traces are the same, across different algorithms, but the average number of distinct locations sent to the LBS is not the same. This is because of the high randomness in the {\em Uniform IID}, {\em Agg Mobility IID}, and {\em RW Agg Mobility} that select fake traces from all possible locations. Our method and the {\em RW User Profile} method have both lower diversity overhead and lower semantic overhead in the set of fake locations. Our method, clearly outperforms all the tested methods, especially the random strategies. For the case of {\em RW User Profile} method, the privacy level against tracking attack gets closer to what we achieve (which is very close to the maximum), due to the fact that the fake traces generated by {\em RW User Profile} are geographically very similar to the true location of the user, and hence creates high confusion, hence error, for the adversary. Note that the {\em RW User Profile} is never a privacy-preserving fake injection method as the adversary can easily de-anonymize and profile the user, no matter if he makes mistakes on exactly tracking the user at each access time (as shown here).

Whereas, our method is ensured to have minimal geographic mutual information with the true trace, thus it is robust against profiling attack. We also ensure that the fake traces have small differential semantic similarity, thus they are robust against de-anonymization. Additionally, the plot shows that our method is the strongest fake generating algorithm against an attacker who is interested in filtering out the fake locations and localize the user over time.

\section{Related Work}\label{sec:relatedwork}

Location obfuscation is a prevalent non-cryptographic technique to protect location privacy. It does not require changing the infrastructure, as it can also be done all on the user's side either by altering (perturbing) the location coordinates to be reported or by sending fake location reports interleaved or along with the true location of the user.

Many location perturbation techniques have been proposed in the literature, usually based on adding some noise to the user's location coordinates or reducing its granularity, e.g., \cite{AndresBCP13, ardagna2011obfuscation, HohGXA07, ShokriTTHB12}. The downside of these techniques is that they reduce the service quality of the user in interaction with the location-based service (LBS) provider. This is because the server provides contextual information related to the shared location and not the true location of the user. So, users have to trade service quality to obtain their required level of privacy. Optimal solutions for location perturbation techniques are proposed \cite{BordenabeCP14, ShokriTTHB12} which show the high cost of location privacy on service quality using perturbation.

Hiding the user's true location among fake locations is a promising yet very little-explored approach to protecting location privacy. There are few simple techniques proposed so far: adding independently selected fake locations drawn from the population's location distribution \cite{ShokriTDHL11}, generating dummy locations at random as a random walk on a grid \cite{KidoYS05, YouPL07}, constructing fake driving trips by building the path between two random locations on the map given the more probable paths traveled by drivers \cite{Krumm09b}, or adding noise to the paths generated by road trip planner algorithms \cite{ChowG09}. These solutions lack a formal model for human mobility and do not consider the semantics associated with sequence of locations visited by people over time. Thus, the generated traces can be distinguished from real location traces.

This paper, to the best of our knowledge, is the first that proposes a systematic methodology for generating fake location traces based on statistical features of both geographical and semantic dimensions of real traces, and based on a metric to measure how realistic a synthetic trace is. Moreover, we introduce multiple privacy tests to ensure that the published/shared fake traces themselves do not leak information about real seed traces. Our evaluation on real data also shows the clear advantage of our algorithm with respect to other existing approaches against known inference attacks.

\section{Conclusions} \label{sec:conclusions}

This is the first paper to address the problem of generating realistic synthetic location traces based on a quantitative metrics. Generating such traces is very useful to protect privacy of users when they share location with location-based services, or when we want to publish a dataset of locations to be used for various research reasons. Based on well-established statistical methods, we propose two metrics to quantify geographic and semantic features of human mobility.  Using these metrics, we propose efficient algorithms to generate fake traces that do not leak geographic information about real individuals (guaranteed using a privacy rejection test), yet highly resemble the mobility of a population semantically. We show that inference attacks cannot identify the true location of mobile users if our fake traces are used as protection. We also quantitatively show that our method is superior to all existing methods of generating fake traces.

    \balance

\end{document}